\documentclass{article}

\usepackage{arxiv}

\usepackage[utf8]{inputenc} 
\usepackage[T1]{fontenc}    
\usepackage{hyperref}       
\usepackage{url}            
\usepackage{booktabs}       
\usepackage{amsfonts}       
\usepackage{nicefrac}       
\usepackage{microtype}      
\usepackage{lipsum}
\usepackage{graphicx}
\usepackage{subcaption}
\usepackage{mathptmx} 
\usepackage{amsmath}
\usepackage{lscape}
\usepackage{pdflscape}
\graphicspath{ {./images/} }

\title{A Real Benchmark Swell Noise Dataset for Performing Seismic Data Denoising via Deep Learning}

\author{
Pablo M. Barros \\
  CENPES, \\
  Petrobras\\
  Rio de Janeiro, Brazil \\
  \texttt{pablobarros@petrobras.com.br} \\  
  \And
Roosevelt de L. Sardinha \\ 
  COPPE \\ 
  Federal University of Rio de Janeiro \\
  Rio de Janeiro, Brazil \\
  \texttt{roosevelt1@gmail.com} \\
  \And
Giovanny A. M. Arboleda \\
  COPPE \\ 
  Federal University of Rio de Janeiro \\
  Rio de Janeiro, Brazil \\
  \texttt{giovannyresearcher@gmail.com} \\
  \And
  Lessandro de S. S. Valente \\
  COPPE \\ Federal University of Rio de Janeiro \\
  Rio de Janeiro, Brazil \\
  \texttt{lessandro.sadala@gmail.com} \\
  \And
  Isabelle R. V. de Melo \\ 
  COPPE \\ Federal University of Rio de Janeiro \\
  Rio de Janeiro, Brazil \\
  \texttt{misabellerv@gmail.com} \\
  \And
  Albino Aveleda \\
  COPPE \\ Federal University of Rio de Janeiro \\
  Rio de Janeiro, Brazil \\
  \texttt{bino@nacad.ufrj.br} \\ 
  \And 
  André Bulcão \\
  CENPES \\ Petrobras \\ Rio de Janeiro, Brazil \\
  \texttt{bulcao@petrobras.com.br} 
  \And
  Sergio L. Netto \\
  COPPE \\ Federal University of Rio de Janeiro \\
  Rio de Janeiro, Brazil \\
  \texttt{sergioln@smt.ufrj.br} \\
  \And
  Alexandre G. Evsukoff \\
  COPPE \\ Federal University of Rio de Janeiro \\
  Rio de Janeiro, Brazil \\
  \texttt{alexandre.evsukoff@coc.ufrj.br} \\
}

\begin{document}
\maketitle
\begin{abstract}
The recent development of deep learning (DL) methods for computer vision has been driven by the creation of open benchmark datasets on which new algorithms can be tested and compared with reproducible results. Although DL methods have many applications in geophysics, few real seismic datasets are available for benchmarking DL models, especially for denoising real data, which is one of the main problems in seismic data processing scenarios in the oil and gas industry. This article presents a benchmark dataset composed of synthetic seismic data corrupted with noise extracted from a filtering process implemented on real data. In this work, a comparison between two well-known DL-based denoising models is conducted on this dataset, which is proposed as a benchmark for accelerating the development of new solutions for seismic data denoising. This work also introduces a new evaluation metric that can capture small variations in model results. The results show that DL models are effective at denoising seismic data, but some issues remain to be solved.
\end{abstract}


\section{Introduction}
\label{sec:1}
Seismic data processing is a primary activity in geosciences; it consumes enormous amounts of human and computational resources, in addition to requiring the acquisition, storage, and processing of large databases. Offshore seismic reflection data are inevitably contaminated by noise derived from several sources, such as sea waves, turbulence along the streamers of seismic cables, wind, and instruments. Noise can compromise the quality of seismic data processing and hinder the further interpretation of geological features, which can have relevant decision-making consequences in exploration and reservoir management scenarios. Therefore, noise attenuation and removal play key roles in the quality of seismic data processing and interpretation strategies. Typically, noise attenuation methods are based on filtering techniques, which presuppose prior noise characteristic knowledge. Noise filtering is an efficient approach when a specific spectral noise characteristic can be easily separated from the observed signal. However, filtering parameter adjustments are often performed manually by domain experts and are time-consuming.

The goal of denoising is to restore the original signal to the greatest extent possible, eliminating the effect of noise and increasing the signal-to-noise ratio (SNR). Image denoising is a classic problem in signal processing \cite{7}, but it remains a challenging inverse problem, and its solution may not be unique. In many applications, particularly in cases involving seismic data, distinguishing signals from noise is difficult, and images may lose some important details and features after the denoising process.

Recently, Deep Learning (DL) methods have produced good image denoising results. \cite{22} presented a review of more than 200 articles on image denoising applications of DL models. The problem of image denoising was divided into four categories according to their noise types: additive white noise, real noise, hybrid noise, and blind image denoising. In the first three categories, a clean image is corrupted with some type of noise and used for model fitting. Different approaches have been developed according to their added noise types and DL model topologies. In the last category, a model is directly fitted to noisy data, and clean images are used for model performance evaluations.

In seismic data denoising cases, most approaches based on DL methods use data with additive white Gaussian noise (AWGN) \cite{1}. AWGN is not a good predictor of seismic noise since seismic data have very different noise sources \cite{23}. Some authors have used passive ambient noise, recorded without sources, for seismic denoising \cite{4, 11}. However, the associated noise files are not available, so their experiments are not reproducible, making model benchmarking difficult.

The recent development of DL methods for computer vision has been driven by the creation of open benchmark datasets on which different algorithms can be tested and compared. Nevertheless, the oil and gas industry still resists making its data available, and few real seismic datasets have been produced for benchmarking DL models in denoising problems. This article introduces a benchmark seismic denoising dataset composed of synthetic seismic data corrupted with real noise extracted after implementing a filtering process. The dataset is proposed as a benchmark for accelerating the development of new solutions for denoising seismic data. The proposed benchmark dataset is used in this work to compare two common DL architectures. This work also introduces an evaluation metric that can capture small variations in model results.

The remainder of the article is organized as follows. Section 2 presents a review of the recent works on DL applied to seismic data denoising. Section 3 presents the new dataset proposed as a benchmark. Section 4 presents the DL methods used for denoising in this work. Section 5 described the denoising results achieved on the proposed dataset and provides a discussion. Section 6 presents the conclusions.

\section{Related Works}
\label{sec:2}
Different DL models have been reported in the literature for performing seismic data denoising. In general, these models are based on image denoising methods that use the supervised learning approach with noisy data as their inputs and clean (noise-free) data as their target. While not exhaustive, this literature review was compiled by searching the most important journals in the field. \cite{1} recently presented a highly complete survey of the available machine learning and DL applications in the seismic data processing domain. They reported that denoising is one of the most common applications of DL in seismic data processing (\cite{1}).

\cite{18} proposed the application of a convolutional autoencoder (CAE)-based U-Net model, which was trained on synthetic seismic data containing additive random (Gaussian and spike) noise. \cite{10} used a residual neural network (ResNet) trained on synthetic data corrupted by AWGN. \cite{11} combined the discrete wavelet transform with several residual neural network (DWT-ResNet) topologies to denoise synthetic and real single-channel (1D) seismic data corrupted with synthetic and real noise. \cite{23} applied convolutional neural networks (CNNs) to remove random noise from synthetic seismic data. \cite{21} presented a CAE and compared it with classic wavelet and f-x deconvolution methods to remove random noise from synthetic seismic data.

\cite{13} used the denoising CNN (DnCNN) model \cite{26} and a variation using the U-Net model as its backbone (called DUDnCNN). The DnCNN model is an image denoising method in which noise is used as the model training target instead of clean data. The predicted noise is subtracted from the noisy data to obtain a prediction for the clean data. The authors used poststack seismic data from the Kaggle salt body segmentation competition with Gaussian noise and reported that the DUDnCNN model performed better than the DnCNN did. \cite{9} used a MobileNet v2 model combined with singular value decomposition (SVD) to remove AWGN from synthetic seismic data.

\cite{14} evaluated different multiscale dilated convolutional network (MDCN) topologies. They simulated a synthetic seismic dataset with passive noise data recorded in a desert area (Talimu Basin) without firing a cannon. Compared with the bandpass filter, f-x filter, and DnCNN models, the proposed model achieved better results.

Generative adversarial networks (GANs) \cite{6} have also been used for seismic denoising. \cite{5} employed a pix2pix model using synthetic seismic data and AWGN. \cite{15} presented a model called a multiscale residual density GAN (MSRD-GAN) for seismic denoising using synthetic seismic data and AWGN. The model was compared with classic approaches such as the wavelet transform and DL techniques such as the standard GAN and DnCNN. The results showed that the MSRD-GAN performed better at different noise levels. \cite{16} presented the SeisGAN model, which simultaneously performs denoising and superresolution on synthetic seismic data corrupted with AWGN. \cite{3} proposed a new model called the progressive denoising network (PDN) for random and coherent seismic data denoising. The PDN model performs layer-by-layer denoising with feature extraction and denoising subnetworks applied to a synthetic dataset. The random passive noise dataset was recorded under no-source conditions, and the coherent noise was simulated. The model produced good results compared with those of the DnCNN and its GAN variation (DnGAN), as well as with those of the bandpass filter and robust principal component analysis (RPCA).
\cite{4} also proposed a dilated CNN (D-CNN) and a gradual denoising strategy for denoising a synthetic seismic dataset corrupted with real passive ambient noise. The noisy dataset was augmented via the Wasserstein GAN (WGAN), which creates synthetic noise data with similar probability distributions. The model was compared with the DnCNN and ensemble empirical mode decomposition (EEMD). \cite{2} proposed a Swin transformer block integrated with a U-Net model for denoising real and synthetic seismic data corrupted with AWGN.

Table \ref{tab:table1} presents a summary of the related works, where most of the reviewed papers used synthetic seismic data and AWGN (or unspecified random noise) in their experiments and mostly differed in terms of their utilized models. This work presents a seismic denoising benchmark dataset consisting of four synthetic seismic subdatasets corrupted with two real swell noise files extracted after implementing a filtering process, as described in the next section.

\begin{table}[htpb]
\centering
\caption{Summary of the current deep learning models for seismic denoising}
\label{tab:table1}
\begin{tabular}{lllp{110pt}}
\hline\noalign{\smallskip}
\textbf{Reference} & \textbf{Model} & \textbf{Seismic Data} & \textbf{Noise Type} \\
\noalign{\smallskip}\hline\noalign{\smallskip}
\cite{10} & ResNet         & Synthetic                           & AWGN \\
\cite{5}  & Pix2pix        & Synthetic                           & AWGN \\
\cite{23} & CNN            & Synthetic                           & Unspecified random noise \\
\cite{18} & U-Net          & Synthetic                           & AWGN and spike noise \\
\cite{21} & CAE            & Synthetic                           & Unspecified random noise \\
\cite{14} & MDCN           & Synthetic                           & Real passive noise \\
\cite{13} & DnCNN          & Poststack                           & AWGN \\
\cite{9}  & MobileNet v2   & Synthetic                           & AWGN \\
\cite{3}  & PDN            & Synthetic                           & Real passive noise and coherent noise \\
\cite{4}  & D-CNN          & Synthetic                           & Real passive noise and coherent noise \\
\cite{11} & DWT-ResNet     & Synthetic and real (single channel) & Real noise and synthetic noise \\
\cite{15} & MSRD-GAN       & Synthetic                           & AWGN \\
\cite{2}  & SSC-UNet       & Synthetic and real                  & AWGN \\
\cite{16} & SeisGAN        & Synthetic                           & AWGN \\
\noalign{\smallskip}\hline
\end{tabular}
\end{table}

\section{Dataset Description}
\label{sec:3}
The synthetic seismic data were generated via numerical simulations using the elastic wave equation and benchmark velocity models to represent clean (noise-free) data. The noisy data, which represent real data, were generated by adding real noise as follows,
\begin{equation}
\label{eq:eq1}
    x = y + \epsilon,
\end{equation}

where $x$ represents noisy data, $y$ is denotes clean data, and $\epsilon$ signifies additive real noise.

Each seismic dataset is presented as a set of shot gathers, each of which is a 2D array in which the lines represent time, and the columns represent the traces or the number of hydrophones in the acquisition cable (cf. Table \ref{tab:table2}). Each trace represents the subsurface reflections recorded by the corresponding hydrophone after a shot. The number of lines for a shot gather depends on the sampling frequency and the total acquisition time.

\begin{figure}[ht]
    \centering
    \begin{subfigure}[b]{0.24\textwidth}
        \centering
        \includegraphics[width=\textwidth]{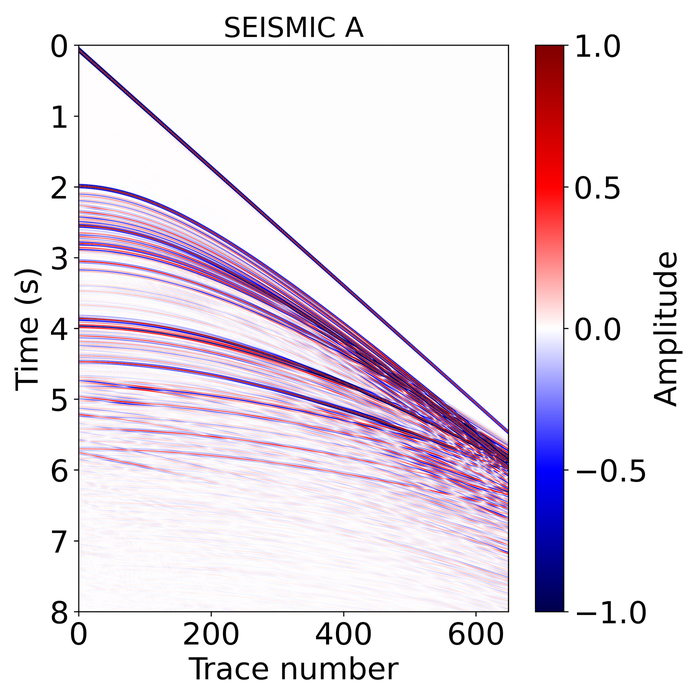}
        \caption{}
        \label{fig:sub-a}
    \end{subfigure}
    \hfill
    \begin{subfigure}[b]{0.24\textwidth}
        \centering
        \includegraphics[width=\textwidth]{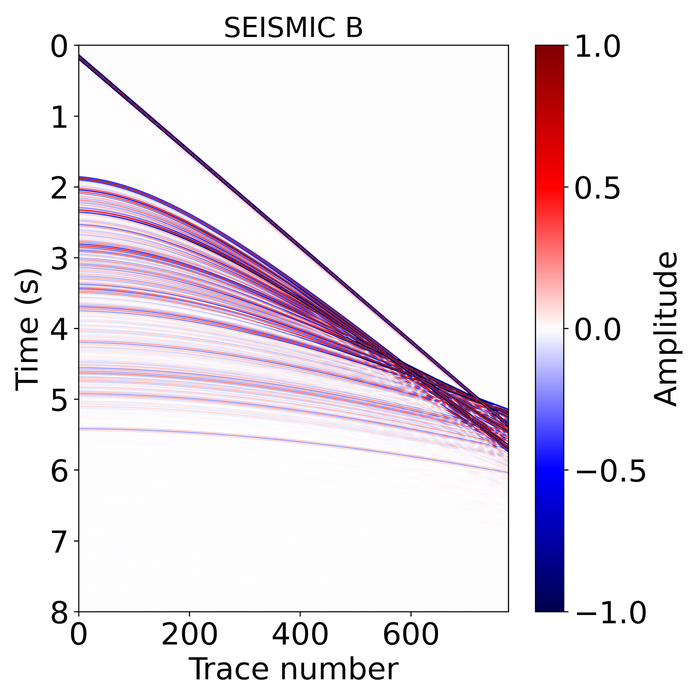}
        \caption{}
        \label{fig:sub-b}
    \end{subfigure}
    \hfill
    \begin{subfigure}[b]{0.24\textwidth}
        \centering
        \includegraphics[width=\textwidth]{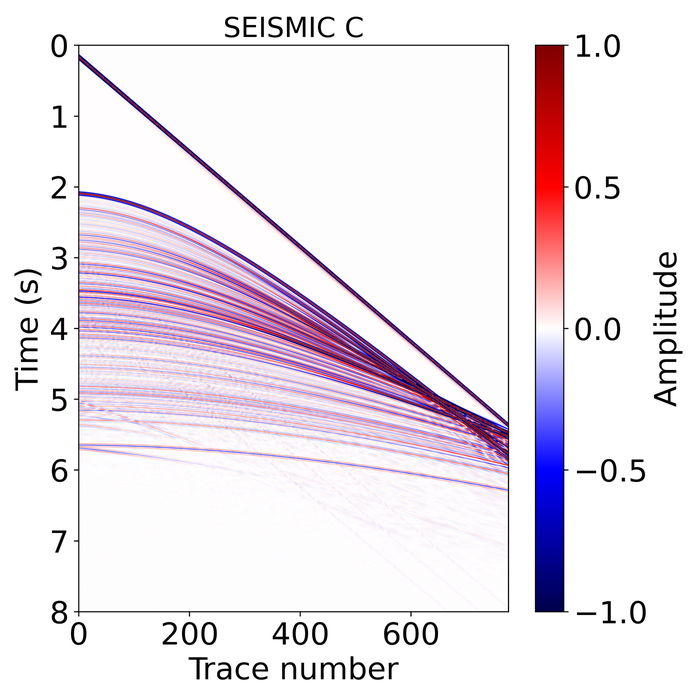}
        \caption{}
        \label{fig:sub-c}
    \end{subfigure}
    \hfill
    \begin{subfigure}[b]{0.24\textwidth}
        \centering
        \includegraphics[width=\textwidth]{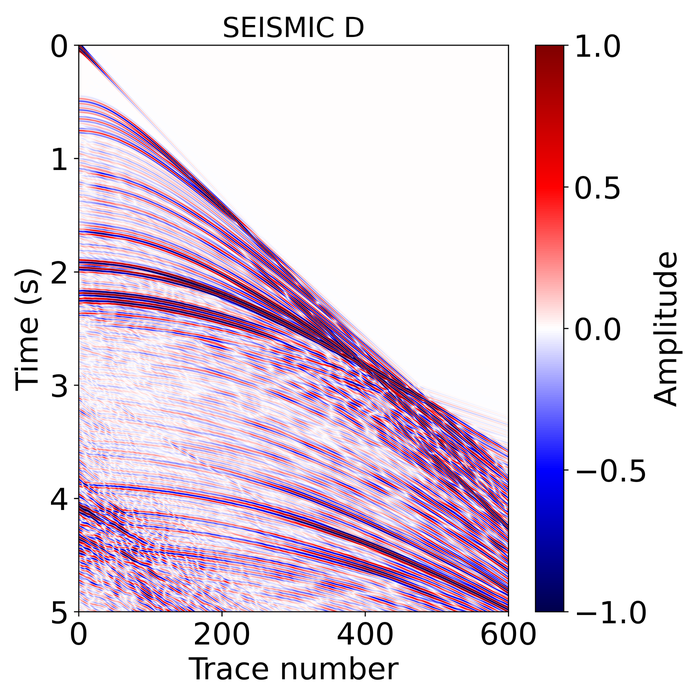}
        \caption{}
        \label{fig:sub-d}
    \end{subfigure}
    \caption{Clean seismic data: (a) Seismic A, (b) Seismic B, (c) Seismic C, and (d) Seismic D.}
    \label{fig:1}
\end{figure}

In this work, two numerical simulations of the seismic responses of two different open-source structural models were used to generate four synthetic datasets, as shown in Table \ref{tab:table2}. The data obtained via the Marmousi structural model differed in terms of their simulation parameters, and the data obtained via the SEAM model were simulated on one inline and one crossline of a 3D model. One seismogram from each synthetic dataset, acquired after performing data rescaling (cf. Sect. \ref{sec:4}), is shown in Fig \ref{fig:1}.

\begin{table}[ht]
\centering
\caption{Synthetic seismic data}
\label{tab:table2}
\begin{tabular}{lllp{50pt}llp{30pt}}
\hline\noalign{\smallskip}
\textbf{FILE} & \textbf{Structural Model} & \textbf{Shot Gathers} & \textbf{Total Number of Traces} & \textbf{Sampling Period} & \textbf{Num. of Samples} & \textbf{Trace Interval} \\
\noalign{\smallskip}\hline\noalign{\smallskip}
SEISMIC A     & MARMOUSI                  & 1,440                 & 933,120                         & 2 ms                     & 4,001                    & 8 s \\
SEISMIC B     & SEAM                      & 1,200                 & 931,200                         & 2 ms                     & 4,001                    & 8 s \\
SEISMIC C     & SEAM                      & 1,200                 & 931,200                         & 2 ms                     & 4,001                    & 8 s \\
SEISMIC D     & MARMOUSI                  & 420                   & 252,000                         & 2 ms                     & 2,501                    & 5 s \\
\noalign{\smallskip}\hline
\end{tabular}
\end{table}

The noise data were obtained from real data by subtracting the processed data from the original raw seismic data (Table \ref{tab:extra1}). Since the noise data came from a different source, they were on a different scale from that of the synthetic data and needed to be rescaled before being added to the clean synthetic data for computing the noisy data, as in Eq. \ref{eq:eq1}.

\begin{table}[htpb]
\centering
\caption{Noise data obtained from real data by subtracting the processed data from the original raw seismic data.}
\label{tab:extra1}
\begin{tabular}{l l c c c c}
\hline
\textbf{FILE} & \textbf{Shot Gathers}        & \multicolumn{1}{l}{\textbf{Total Number of Traces}} & \multicolumn{1}{l}{\textbf{Sampling Period}} & \multicolumn{1}{l}{\textbf{Num. of Samples}} & \multicolumn{1}{l}{\textbf{Trace Interval}} \\ \hline
NOISE 1       & 1904 & 6168956                                              & 2 ms                                          & 4201                                          & 8.4 s                                        \\ 
NOISE 2       & 1672                         & 9630720                                              & 2 ms                                          & 3267                                          & 6.5 s                                        \\ \hline
\end{tabular}
\end{table}

Noise rescaling was performed such that the desired SNR was obtained from the noisy data (clean synthetic data + real noise). For example, consider the desired SNR of the noisy data to be the same as the SNRe of the real data, which is written as 

\begin{equation}
    \textnormal{SNR}_{e} = \dfrac{\textnormal{RMS}(y_{R})}{\textnormal{RMS}(e_{R})},
\end{equation}

where $y_{R}$ is the real clean data, $e_{R}$ is the real noise at the original scale, and $\textnormal{RMS}$ is the root mean square of the signal, which represents the signal energy.

The normalized noise data were computed as

\begin{equation}
    \hat{e_{R}} = \dfrac{e_{R}}{\textnormal{RMS}(e_{R})},
\end{equation}

such that $\textnormal{RMS}(e_{R})=1$.

The noise to be added to the clean synthetic data was the normalized real noise, scaled in such a way that the synthetic noisy data retained the original SNR,

\begin{equation}
\label{eq:eq4}
   \epsilon =  \hat{e_{R}} \dfrac{\textnormal{RMS}(y)}{\textnormal{SNR}_{e}},
\end{equation}

where $y$ represents the clean synthetic data (cf. Eq. \ref{eq:eq1}). It is possible to verify the scale correction effect by calculating the SNR of the noisy data,

\begin{equation}
    \textnormal{SNR} = \dfrac{\textnormal{RMS}(y)}{\textnormal{RMS}(\epsilon)} = \dfrac{\textnormal{RMS}(y)}{\textnormal{RMS}(\hat{e_{R}}) \dfrac{\textnormal{RMS}(y)}{\textnormal{SNR}_{e}}} = \textnormal{SNR}_{e}.
\end{equation}

\begin{figure}[ht]
    \centering
    \begin{subfigure}[b]{0.24\textwidth}
        \includegraphics[width=\textwidth]{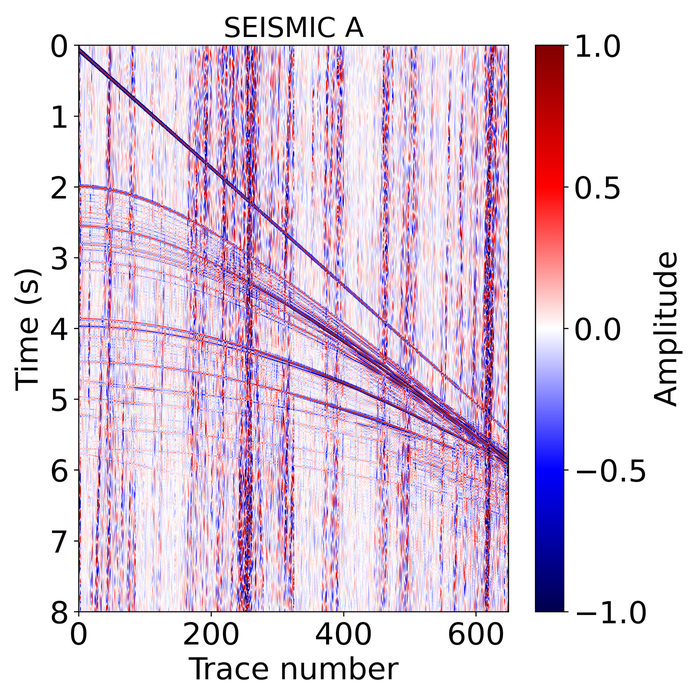}
    \end{subfigure}
    \hfill
    \begin{subfigure}[b]{0.24\textwidth}
        \includegraphics[width=\textwidth]{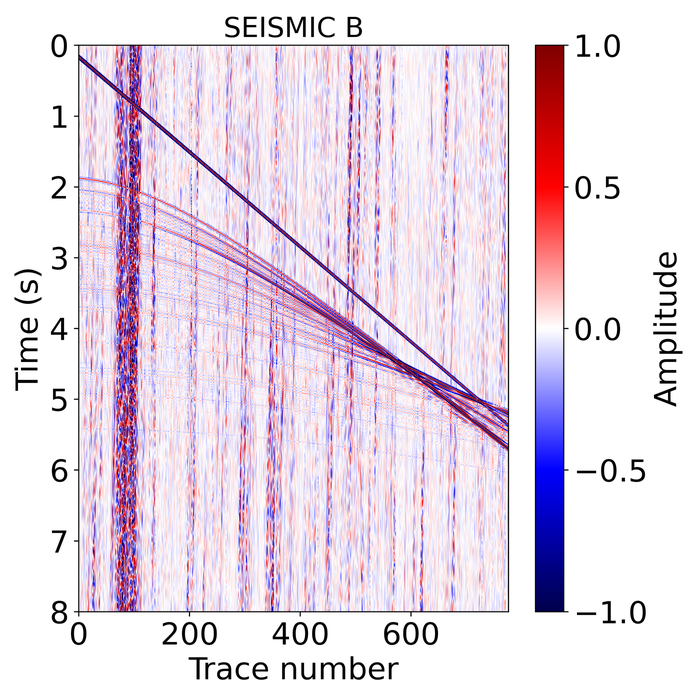}
    \end{subfigure}
    \hfill
    \begin{subfigure}[b]{0.24\textwidth}
        \includegraphics[width=\textwidth]{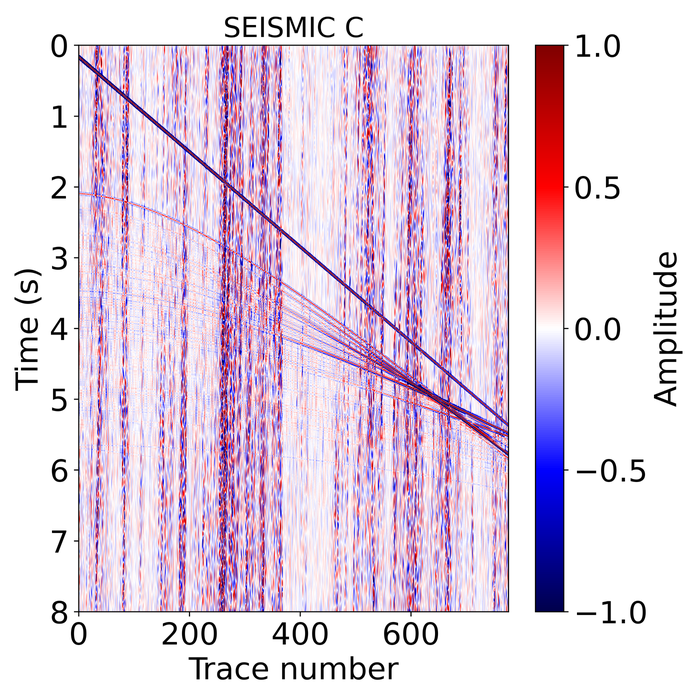}
    \end{subfigure}
    \hfill
    \begin{subfigure}[b]{0.24\textwidth}
        \includegraphics[width=\textwidth]{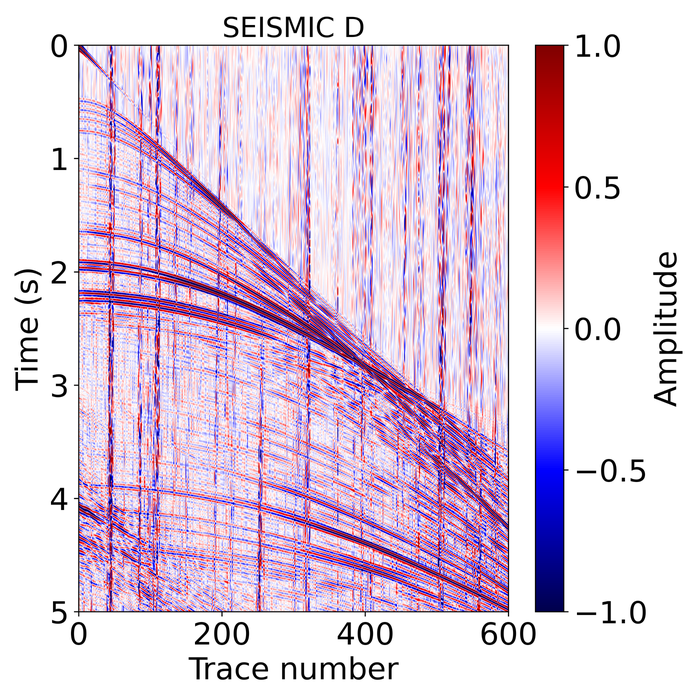}
    \end{subfigure}
    
    \vskip\baselineskip
    
    \begin{subfigure}[b]{0.24\textwidth}
        \includegraphics[width=\textwidth]{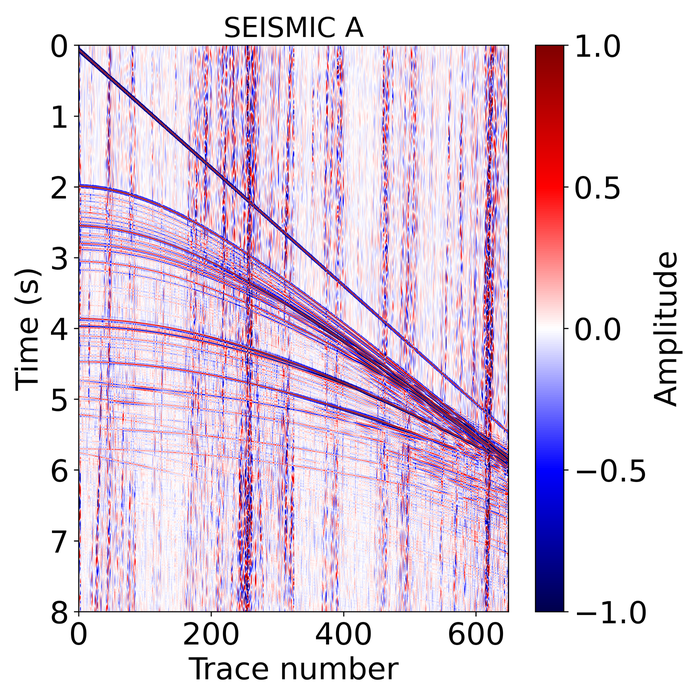}
    \end{subfigure}
    \hfill
    \begin{subfigure}[b]{0.24\textwidth}
        \includegraphics[width=\textwidth]{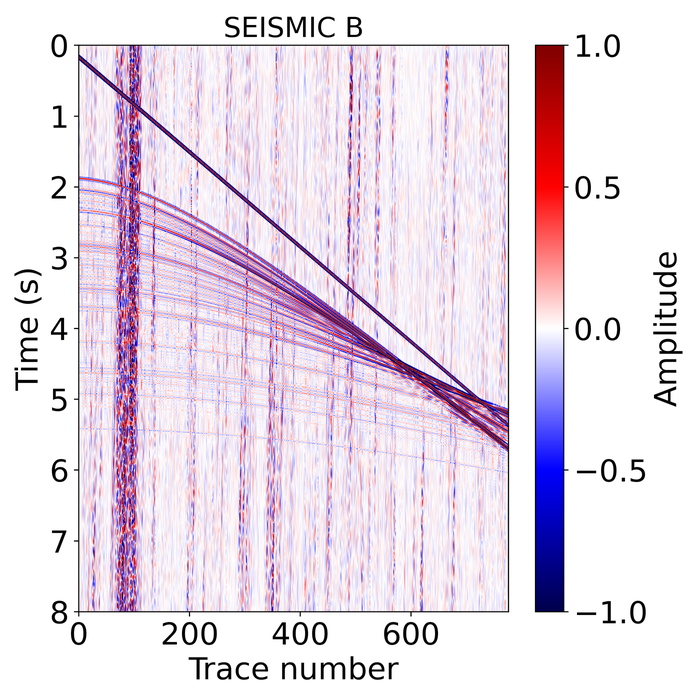}
    \end{subfigure}
    \hfill
    \begin{subfigure}[b]{0.24\textwidth}
        \includegraphics[width=\textwidth]{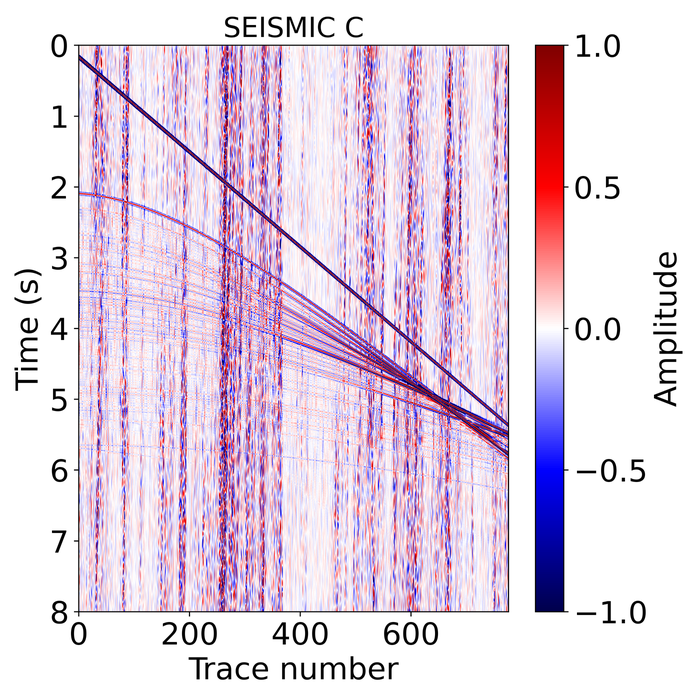}
    \end{subfigure}
    \hfill
    \begin{subfigure}[b]{0.24\textwidth}
        \includegraphics[width=\textwidth]{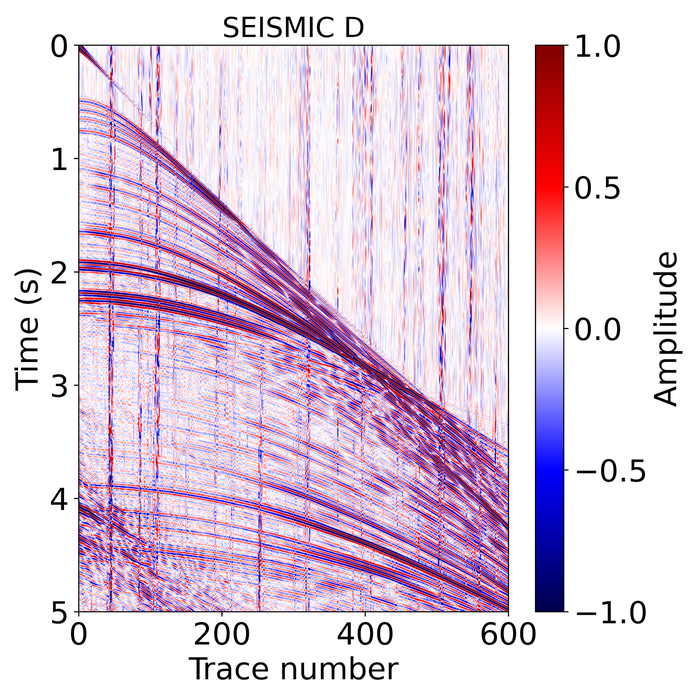}
    \end{subfigure}
    
    \vskip\baselineskip
    
    \begin{subfigure}[b]{0.24\textwidth}
        \includegraphics[width=\textwidth]{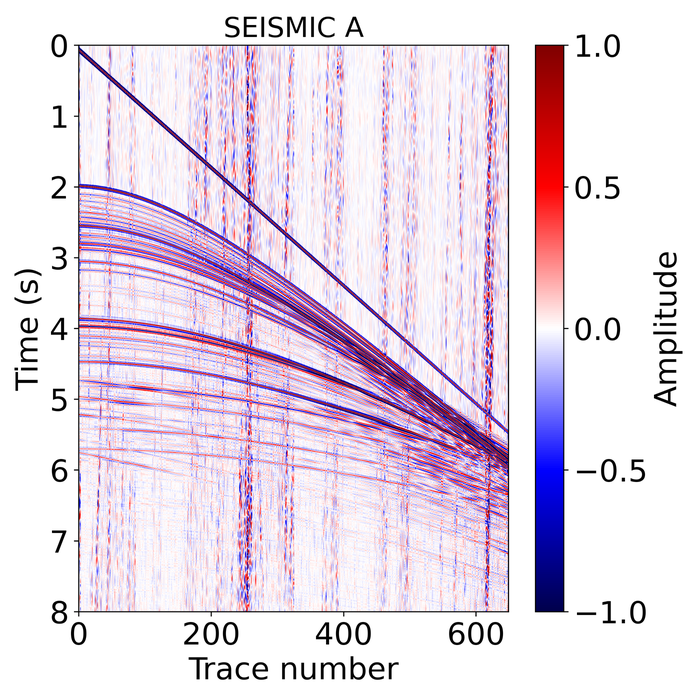}
    \end{subfigure}
    \hfill
    \begin{subfigure}[b]{0.24\textwidth}
        \includegraphics[width=\textwidth]{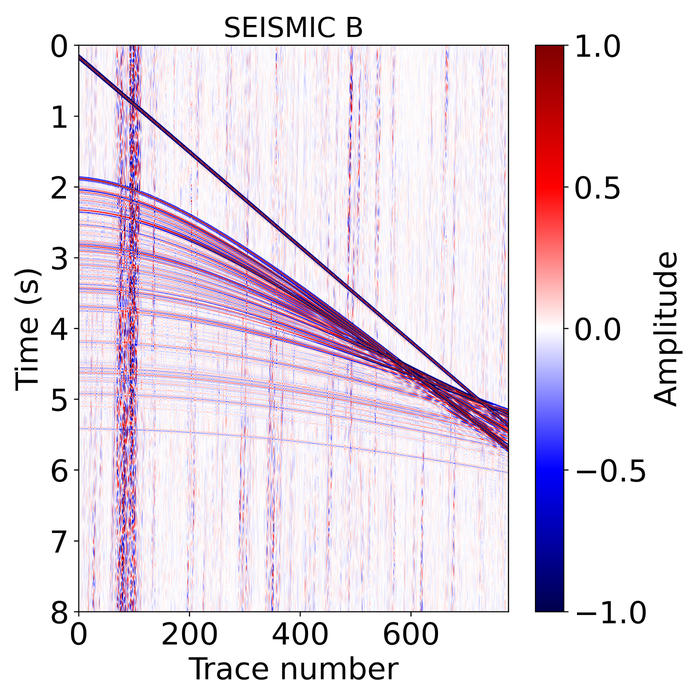}
    \end{subfigure}
    \hfill
    \begin{subfigure}[b]{0.24\textwidth}
        \includegraphics[width=\textwidth]{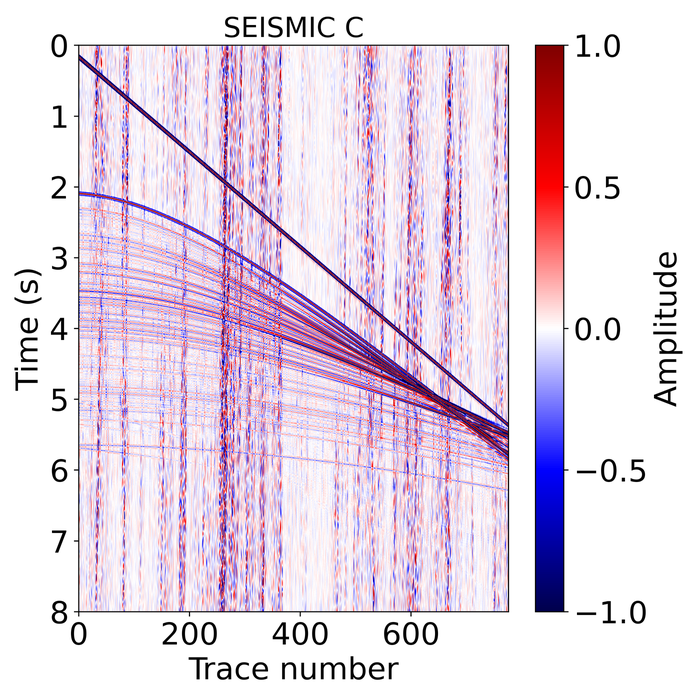}
    \end{subfigure}
    \hfill
    \begin{subfigure}[b]{0.24\textwidth}
        \includegraphics[width=\textwidth]{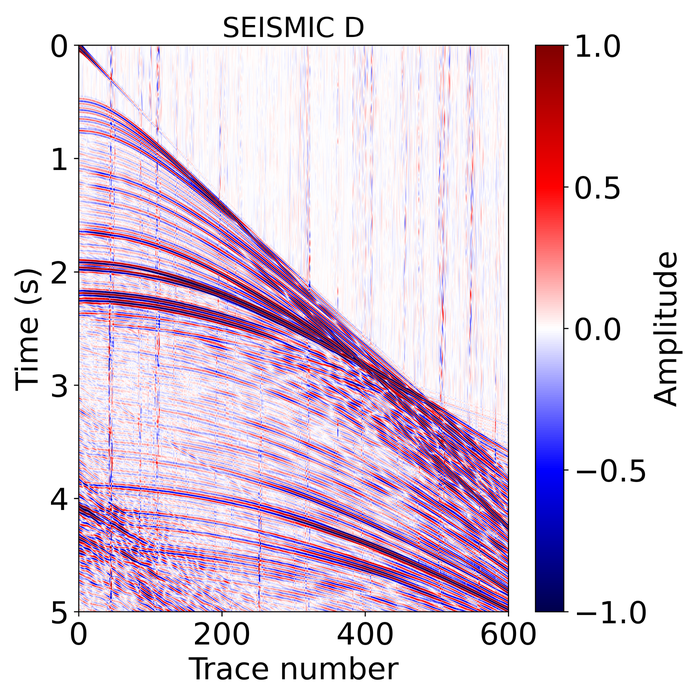}
    \end{subfigure}
    
    \vskip\baselineskip
    
    \begin{subfigure}[b]{0.24\textwidth}
        \includegraphics[width=\textwidth]{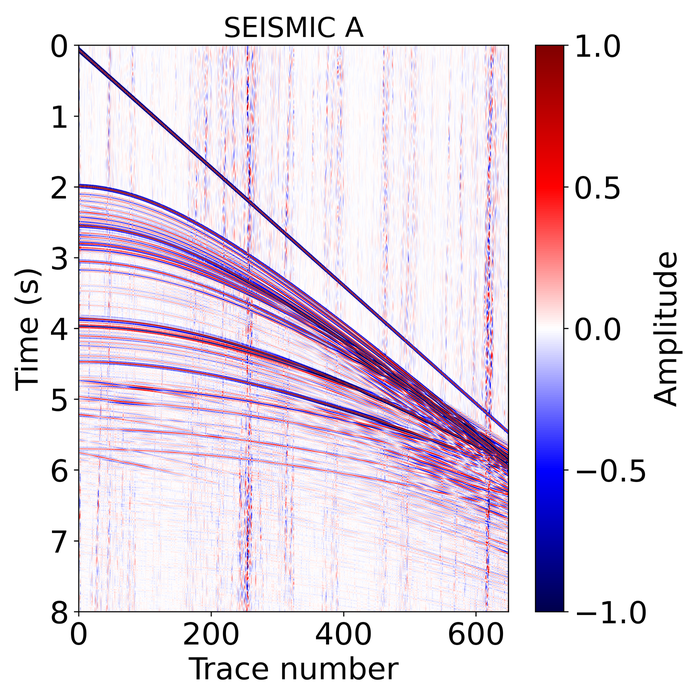}
    \end{subfigure}
    \hfill
    \begin{subfigure}[b]{0.24\textwidth}
        \includegraphics[width=\textwidth]{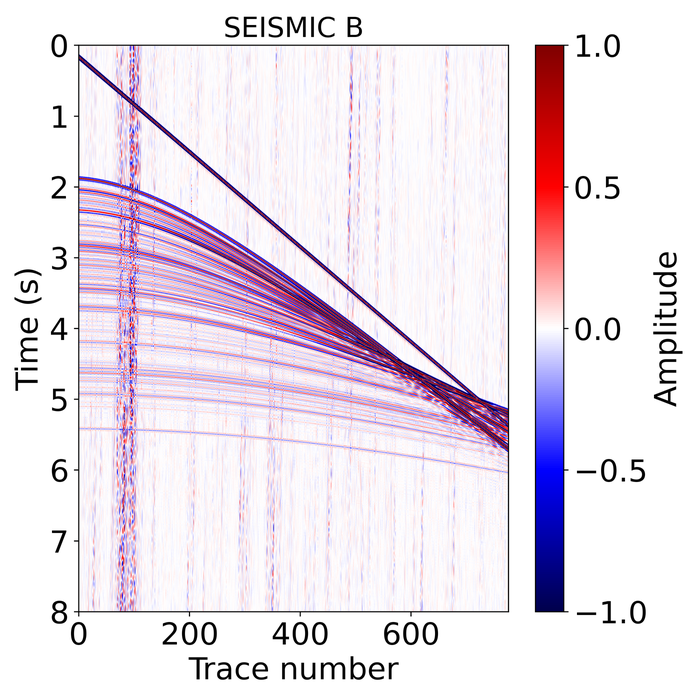}
    \end{subfigure}
    \hfill
    \begin{subfigure}[b]{0.24\textwidth}
        \includegraphics[width=\textwidth]{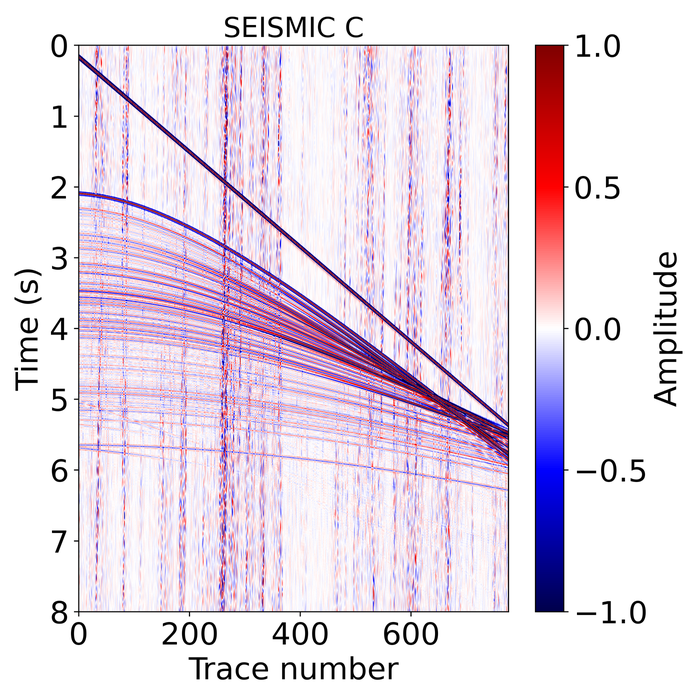}
    \end{subfigure}
    \hfill
    \begin{subfigure}[b]{0.24\textwidth}
        \includegraphics[width=\textwidth]{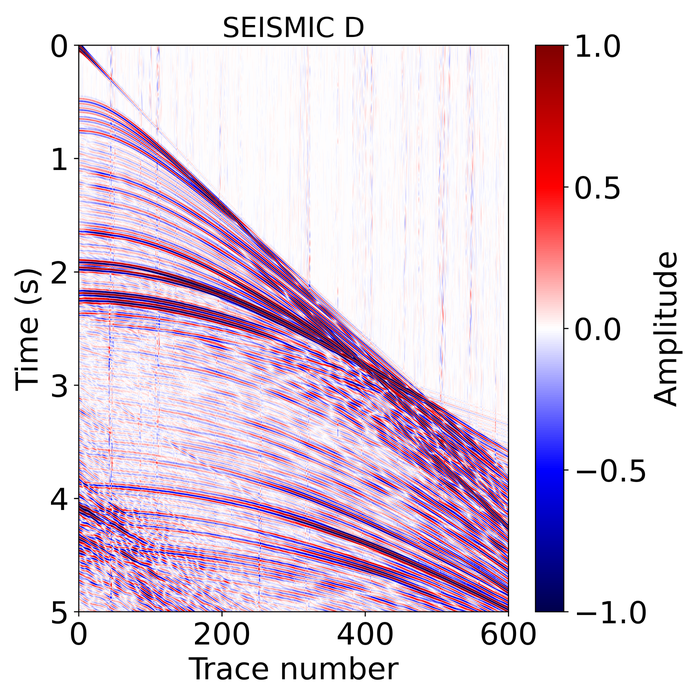}
    \end{subfigure}

    \caption{Noisy seismic data corrupted with NOISE 1: the columns represent the synthetic data (SEISMIC A, B, C and D), and the rows represent the noise levels (L1, L2, L5 and L10).}
    \label{fig:2}
\end{figure}

\begin{figure}[ht]
    \centering
    \begin{subfigure}[b]{0.24\textwidth}
        \includegraphics[width=\textwidth]{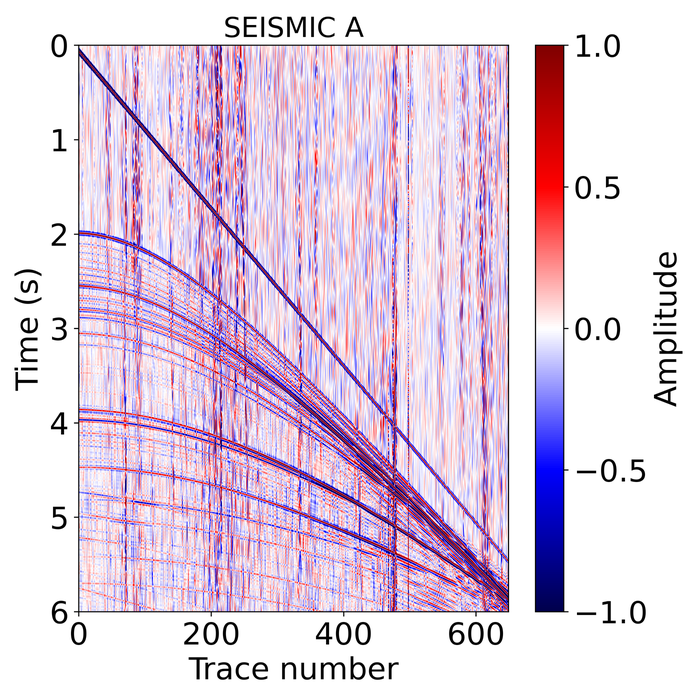}
    \end{subfigure}
    \hfill
    \begin{subfigure}[b]{0.24\textwidth}
        \includegraphics[width=\textwidth]{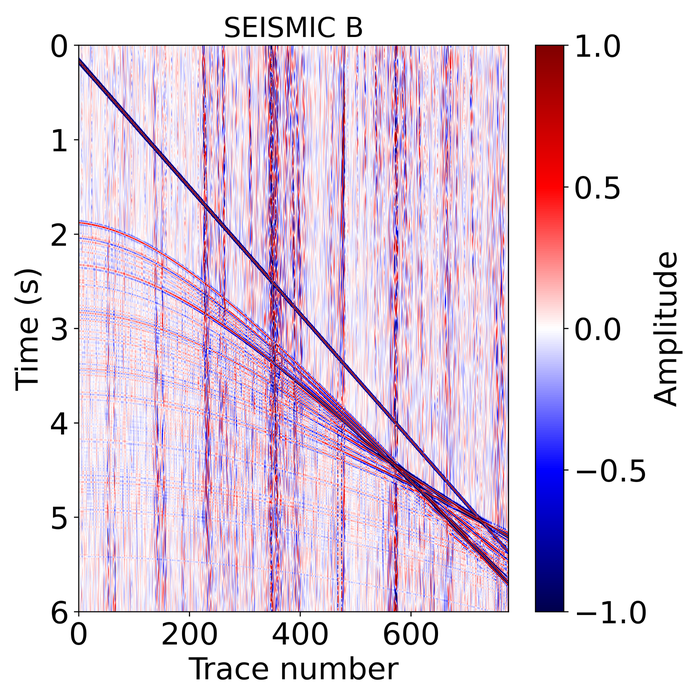}
    \end{subfigure}
    \hfill
    \begin{subfigure}[b]{0.24\textwidth}
        \includegraphics[width=\textwidth]{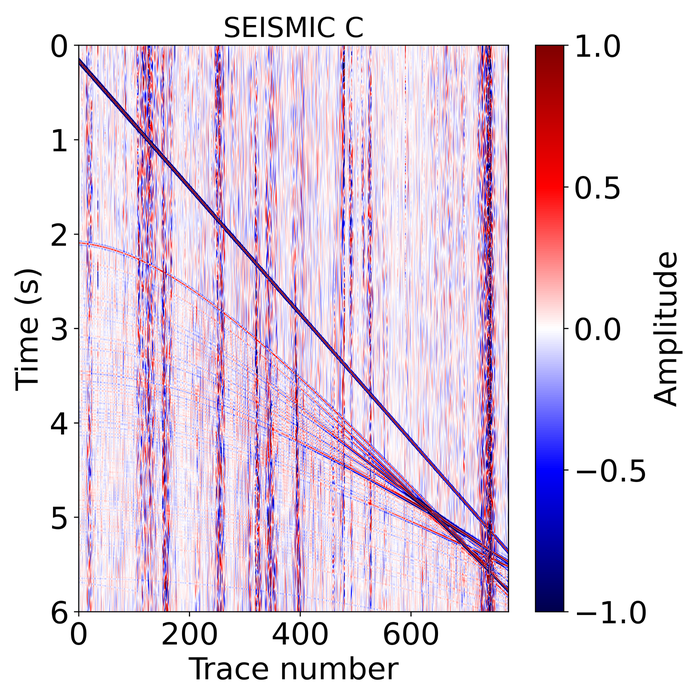}
    \end{subfigure}
    \hfill
    \begin{subfigure}[b]{0.24\textwidth}
        \includegraphics[width=\textwidth]{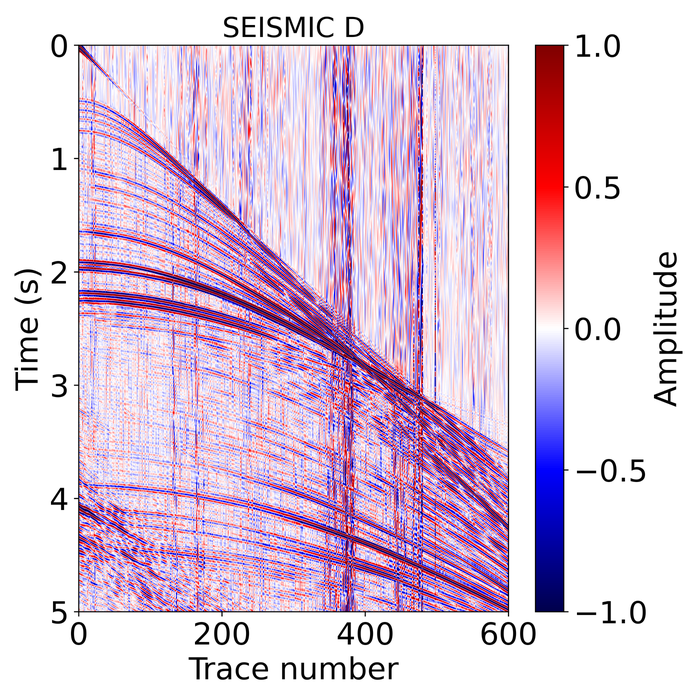}
    \end{subfigure}
    
    \vskip\baselineskip
    
    \begin{subfigure}[b]{0.24\textwidth}
        \includegraphics[width=\textwidth]{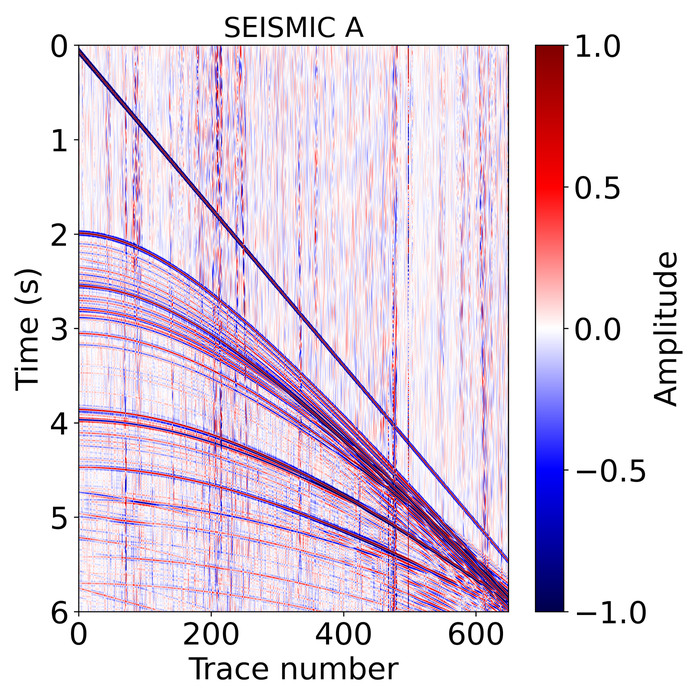}
    \end{subfigure}
    \hfill
    \begin{subfigure}[b]{0.24\textwidth}
        \includegraphics[width=\textwidth]{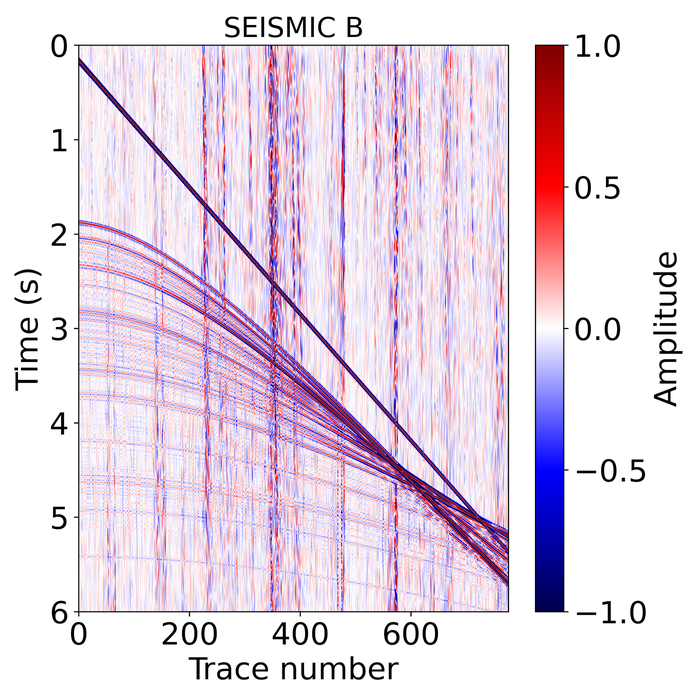}
    \end{subfigure}
    \hfill
    \begin{subfigure}[b]{0.24\textwidth}
        \includegraphics[width=\textwidth]{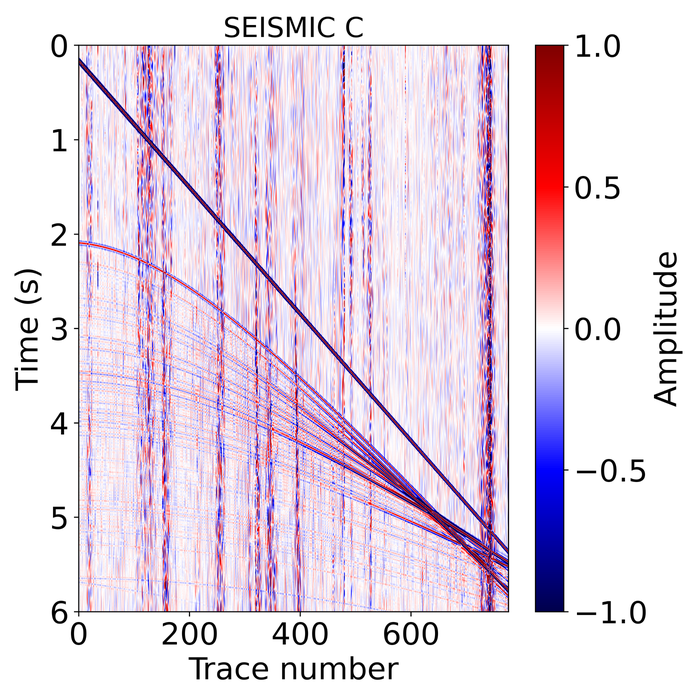}
    \end{subfigure}
    \hfill
    \begin{subfigure}[b]{0.24\textwidth}
        \includegraphics[width=\textwidth]{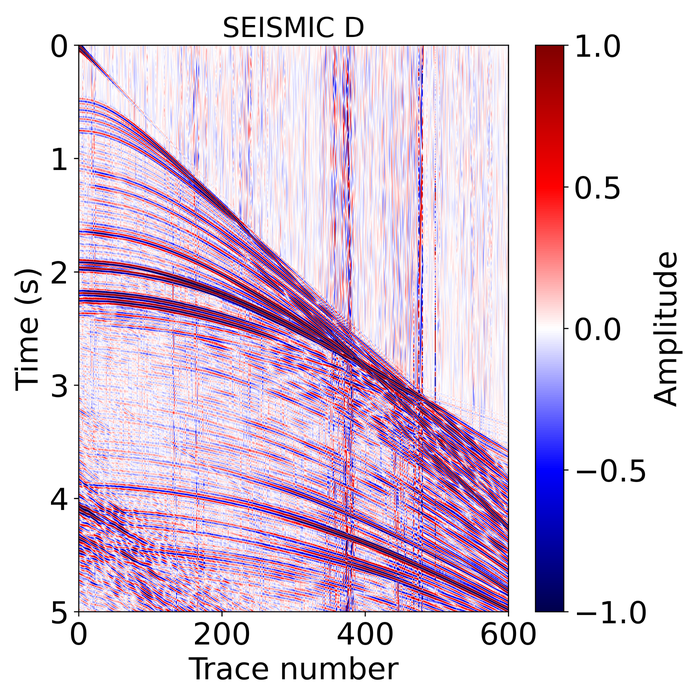}
    \end{subfigure}
    
    \vskip\baselineskip
    
    \begin{subfigure}[b]{0.24\textwidth}
        \includegraphics[width=\textwidth]{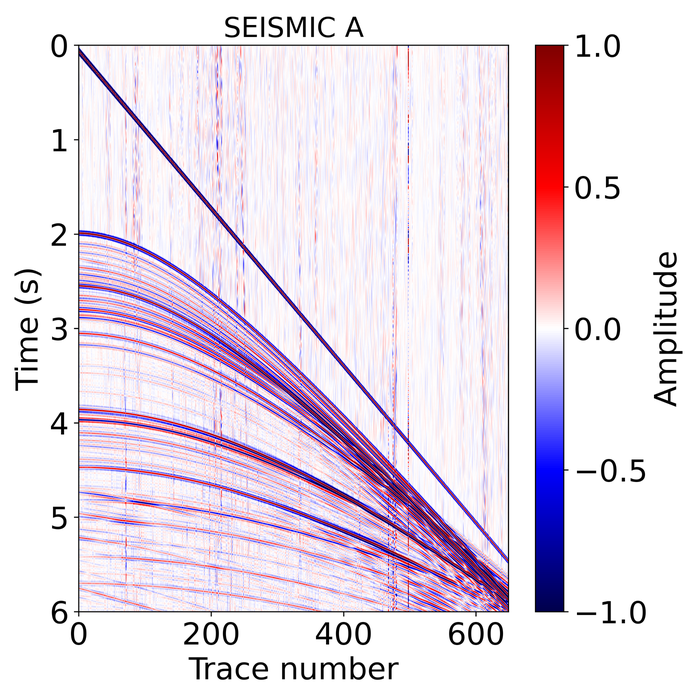}
    \end{subfigure}
    \hfill
    \begin{subfigure}[b]{0.24\textwidth}
        \includegraphics[width=\textwidth]{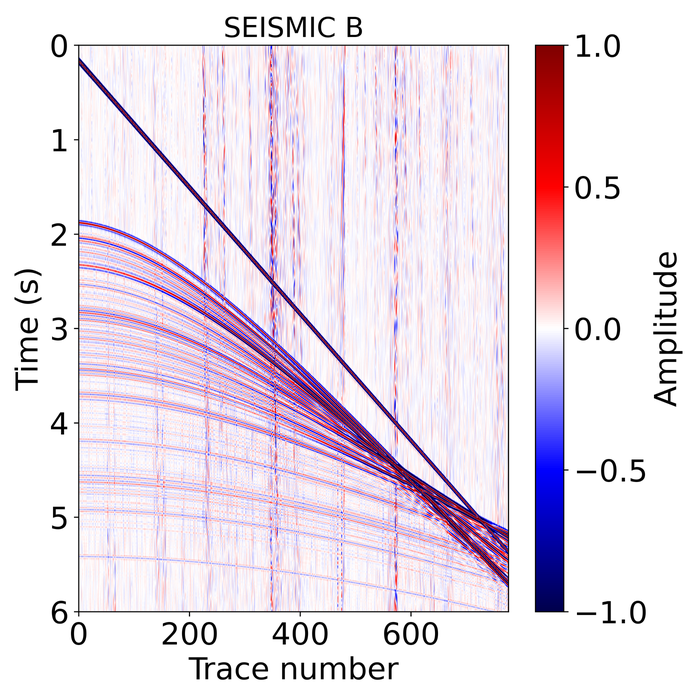}
    \end{subfigure}
    \hfill
    \begin{subfigure}[b]{0.24\textwidth}
        \includegraphics[width=\textwidth]{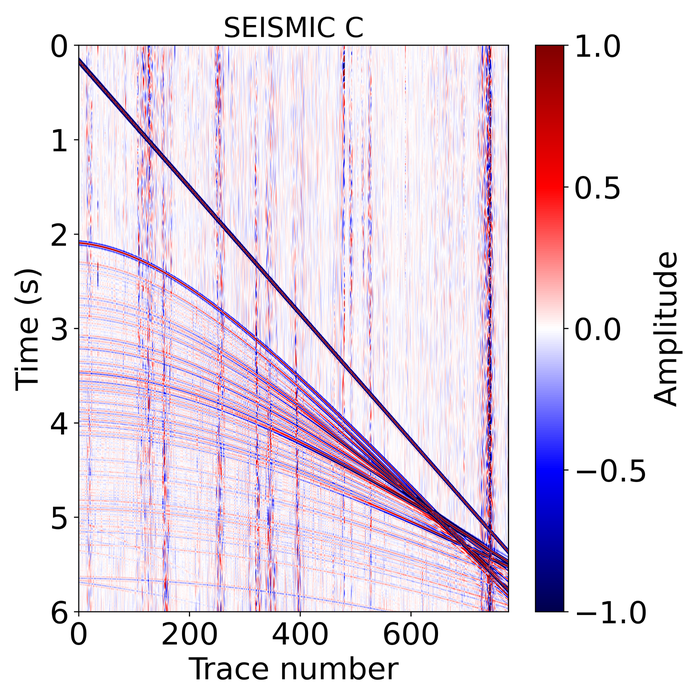}
    \end{subfigure}
    \hfill
    \begin{subfigure}[b]{0.24\textwidth}
        \includegraphics[width=\textwidth]{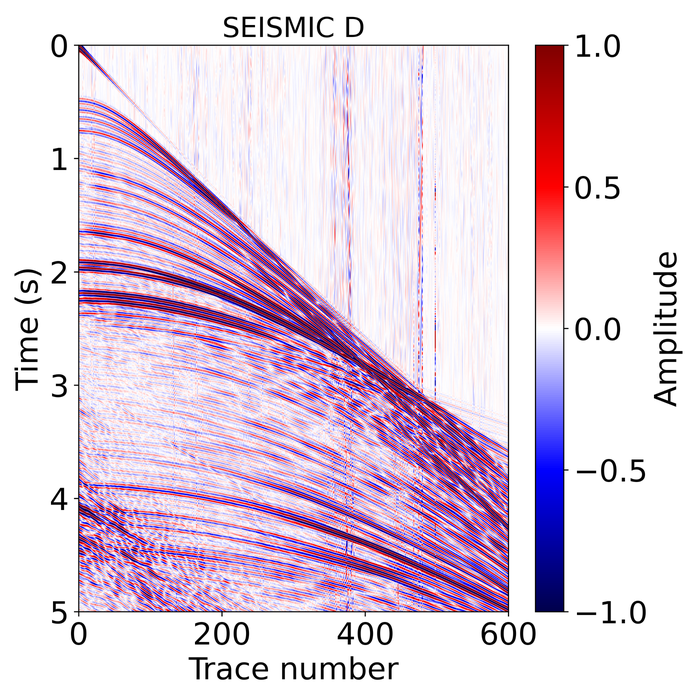}
    \end{subfigure}
    
    \vskip\baselineskip
    
    \begin{subfigure}[b]{0.24\textwidth}
        \includegraphics[width=\textwidth]{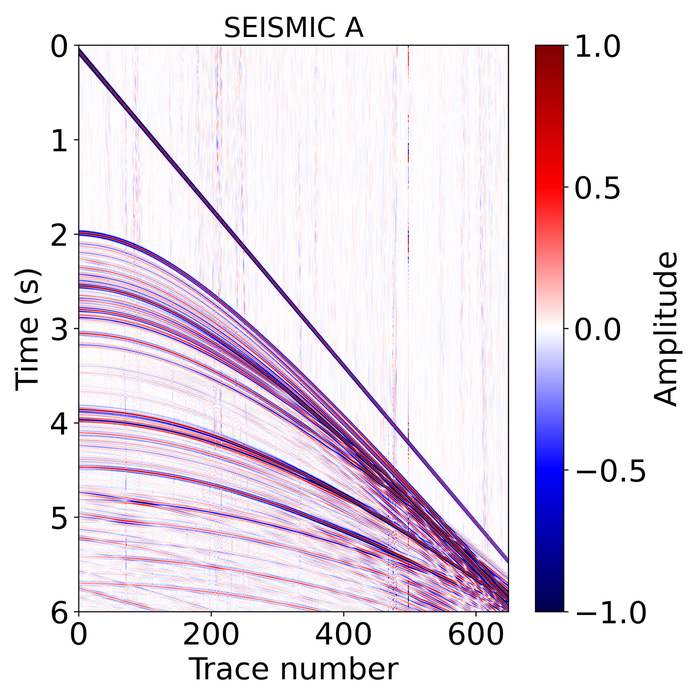}
    \end{subfigure}
    \hfill
    \begin{subfigure}[b]{0.24\textwidth}
        \includegraphics[width=\textwidth]{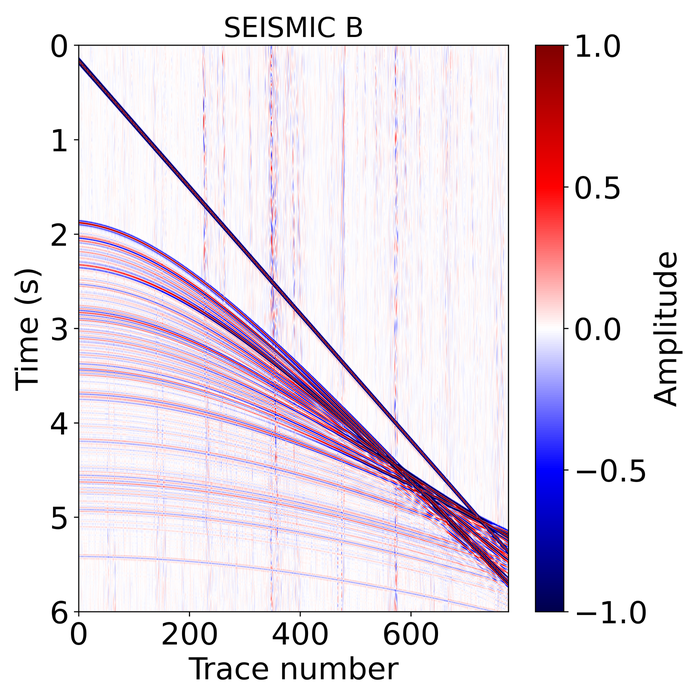}
    \end{subfigure}
    \hfill
    \begin{subfigure}[b]{0.24\textwidth}
        \includegraphics[width=\textwidth]{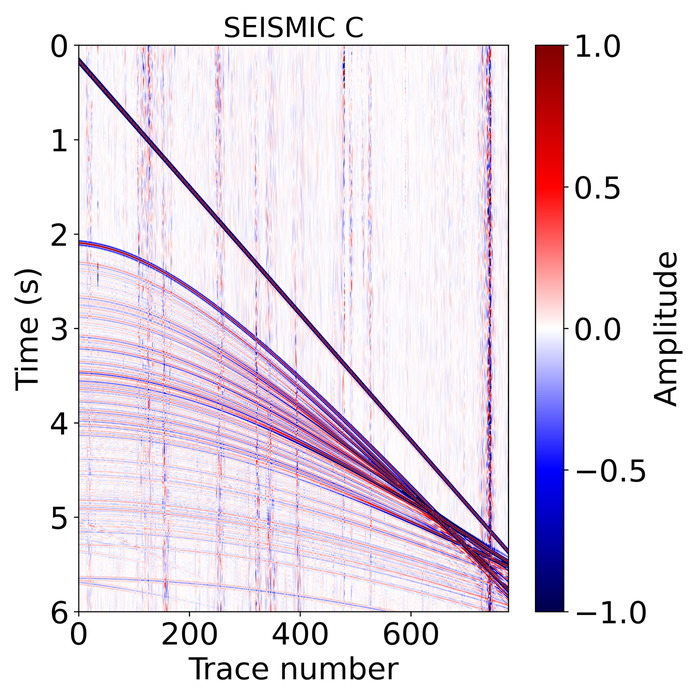}
    \end{subfigure}
    \hfill
    \begin{subfigure}[b]{0.24\textwidth}
        \includegraphics[width=\textwidth]{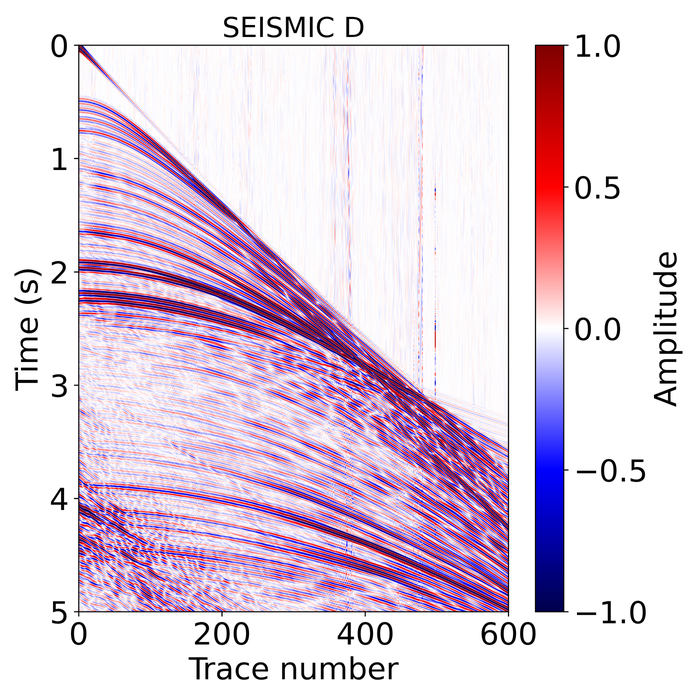}
    \end{subfigure}

    \caption{Noisy seismic data corrupted with NOISE 2: the columns represent the synthetic data (SEISMIC A, B, C and D), and the rows represent the noise levels (L1, L2, L5 and L10).}
    \label{fig:3}
\end{figure}

In this work, two real swell noise files (NOISE 1 and NOISE 2), recorded in different acquisition conditions, were added to the four synthetic seismic datasets (cf. Table \ref{tab:table2}) according to Eq \ref{eq:eq1}. The noise files were rescaled at four different levels according to the desired SNR of the noisy data (cf. Eq \ref{eq:eq4}), L1, L2, L5 and L10, resulting in 16 noisy data points for each noise file. Fig \ref{fig:2} shows the four seismograms of the synthetic seismic data (columns) corrupted with the NOISE 1 file at four different levels (lines). Figure \ref{fig:3} shows the seismic data corrupted with the NOISE 2 file in the same manner. The f-k spectra of these data are shown in Figs. \ref{fig:A1} and \ref{fig:A2}, respectively, in the appendix.

The synthetic data had different trace intervals (cf. Table \ref{tab:table2}), which remained the same when combined with the NOISE 1 file. However, the NOISE 2 file was shorter, so the noisy data had to be cut to approximately 6 s when combined with the synthetic data in SEISMIC A, SEISMIC B and SEISMIC C to directly match the length of the noise file, thus avoiding resampling.

\section{Methods}
\label{sec:4}

\begin{figure}[h!]
  \centering
  \includegraphics[width=0.6\textwidth]{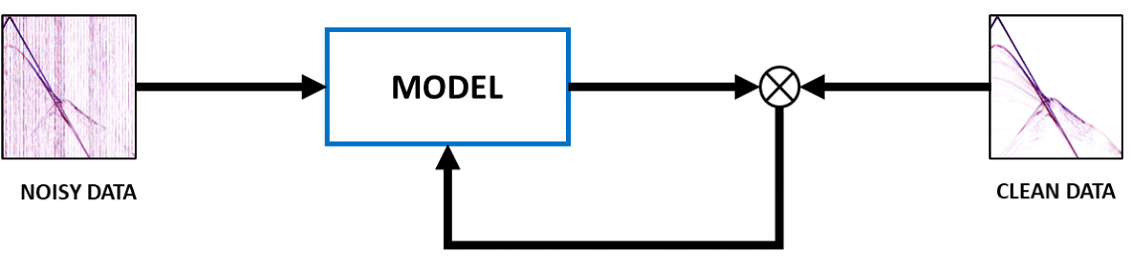} 
  \caption{Supervised DL approach for denoising.}
  \label{fig:4}
\end{figure}

The most common DL approach for denoising is supervised learning, where noisy data x are presented to the utilized model as inputs and clean data y are used as the target, as shown in Fig. \ref{fig:4}. The model is trained to estimate the target as

\begin{equation}
    \hat{y} = f(x,\theta),
\end{equation}

where $\hat{y}$ is the model estimate of the target, x represents the noisy data computed via Eq. \ref{eq:eq1}, and  $\theta$ denotes the model parameters.

Seismic data may be quite different in scale and hence should be rescaled before being processed by the model. In this work, linear scaling was applied to each shot gather of the input seismic data in the form of a modified version of the z score, 
\begin{equation}
\label{eq:7}
    \hat{x} = \dfrac{x - \mu}{3\sigma},
\end{equation}

where $\mu$ is the average of all the values contained in the seismic data and $\sigma$ is the standard deviation. In this scaling scheme, if  is issued from a Gaussian distribution, $99\%$ of the scaled variable $\hat{x} \in \left[  -1, 1  \right]$. Each shot gather has topological neighborhood and texture properties that can be exploited by the convolutional operators in DL models. In this work, DL image denoising methods were applied to the seismic data to identify the coherent information between the input (noisy) data and the target (clean) data.

After performing rescaling, the input and output data of each shot gather were cropped into a set of input‒output pairs of crops with sizes of $256 \times 64$, which were then used to train the model. Preliminary tests were executed using crops with different sizes, but no significant differences were detected. The training set size was specified by the total number of records, where each record was an input‒output pair cropped at the same position. For each shot gather, the associated crops were randomly selected until the required number of records for the training set was reached. Moreover, the crop selection function was constrained to sample the region with the seismic reflections of the shot gather, in other words, the area below the direct wave. All the models were fully convolutional, such that the whole shot gather was presented to each model during the inference phase. In the next sections, the DL models used in this work are presented.

\subsection{FCNN}

\begin{figure}[pb]
  \centering
  \includegraphics[width=0.9\textwidth]{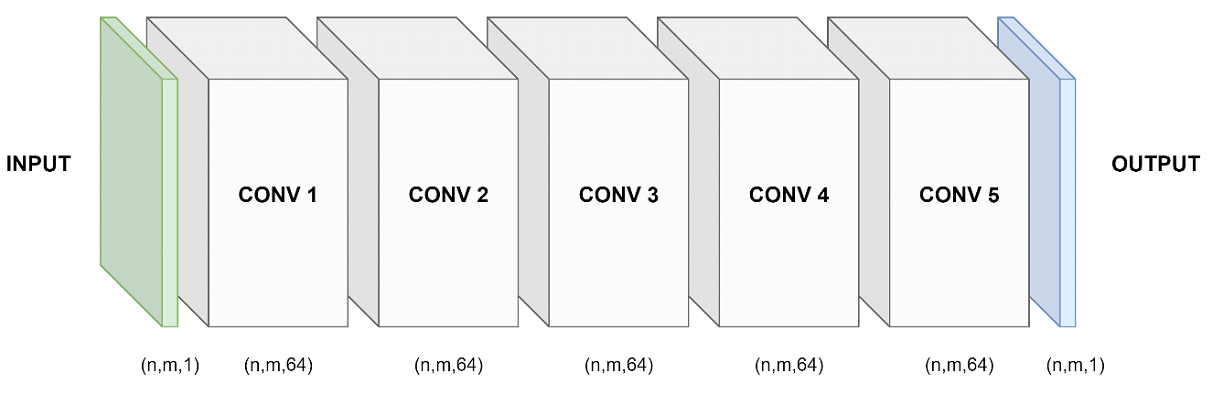} 
  \caption{FCNN model with 5 layers.}
  \label{fig:5}
\end{figure}

The fully CNN (FCNN) is a type of CNN in which a series of convolutional layers are applied to the input \cite{17}. One of the main features of an FCNN is that the model is independent of the input size; thus, the size of the output depends only on the size of the input. This is a useful feature when addressing problems where the inputs may have variable sizes. The 5-layer FCNN model (FCNN-5) employed in this work is presented in Fig. \ref{fig:5}, where only convolutional layer blocks were used, and all layers had the same size as that of the input crop. The model input in the training phase was a crop, and the whole shot gather was presented to the model in the inference phase. Owing to its simplicity, the network training time was usually shorter than that of more complex architectures.

The FCNN-5 model parameters were adjusted via the $L_{1}$ loss function, which yielded better results in preliminary tests,

\begin{equation}
    L_{1} = \dfrac{1}{Nm} \sum_{i=1}^{N} \sum_{t \in c_{i}} |y(t) - \hat{y(t)}|,
\end{equation}

where $y(t)$ is the target value, and $\hat{y(t)}$ is the model prediction obtained for each value of the crop $c_{i}$, $m=16,384$ is the number of values in each $256\times64$ crop, and $N$ is the total number of records included in the training dataset.

The topological details of FCNN-5 are shown in Table \ref{tab:table3}, where the first column represents the layer; the second column denotes the kernel size; the third column represents the output format as a vector [number of channels, crop heig, crop width]; and the last column is the number of parameters in the layer. All the convolutional layers use the parameterized rectified linear unit (PReLU) activation function, in which the slope of the linear part of the function is adjustable by a parameter. The output layer uses 64 $1\times1$ filters and a linear activation function. The model has a total of 1,332,678 trainable parameters.

\begin{table}[pb]
\centering
\caption{FCNN-5 Model Topology}
\label{tab:table3}
\begin{tabular}{lllp{60pt}}
\hline\noalign{\smallskip}
\textbf{Layer (Type)} & \textbf{Filter} & \textbf{Output Shape} & \textbf{Parameters} \\
\noalign{\smallskip}\hline\noalign{\smallskip}
Conv2d-1  & $9\times9$  & {[}64, 256, 64{]} & 5,248 \\
PReLU-2   & -           & {[}64, 256, 64{]} & 1     \\
Conv2d-3  & $9\times9$  & {[}64, 256, 64{]} & 331,840 \\
PReLU-4   & -           & {[}64, 256, 64{]} & 1     \\
Conv2d-5  & $9\times9$  & {[}64, 256, 64{]} & 331,840 \\
PReLU-6   & -           & {[}64, 256, 64{]} & 1     \\
Conv2d-7  & $9\times9$  & {[}64, 256, 64{]} & 331,840 \\
PReLU-8   & -           & {[}64, 256, 64{]} & 1     \\
Conv2d-9  & $9\times9$  & {[}64, 256, 64{]} & 331,840 \\
PReLU-10  & -           & {[}64, 256, 64{]} & 1     \\
Conv2d-11 & $1\times1$  & {[}1, 256, 64{]}  & 65    \\
\noalign{\smallskip}\hline
\end{tabular}
\end{table}

The FCNN models were trained using  records (crops) in the training data, with a batch size of 2048 records. The optimizer was stochastic gradient descent (SGD) with an initial learning rate of 0.01, and early stopping was implemented with a patience term of 50 epochs. The training process was halted if the maximum number of epochs (500) was reached or if the processing time exceeded 10 days. Other optimizers were evaluated, with no significant performance differences exhibited by this model.

\subsection{SRGAN}

\begin{figure}[ht]
  \centering
  \includegraphics[width=1.0\textwidth]{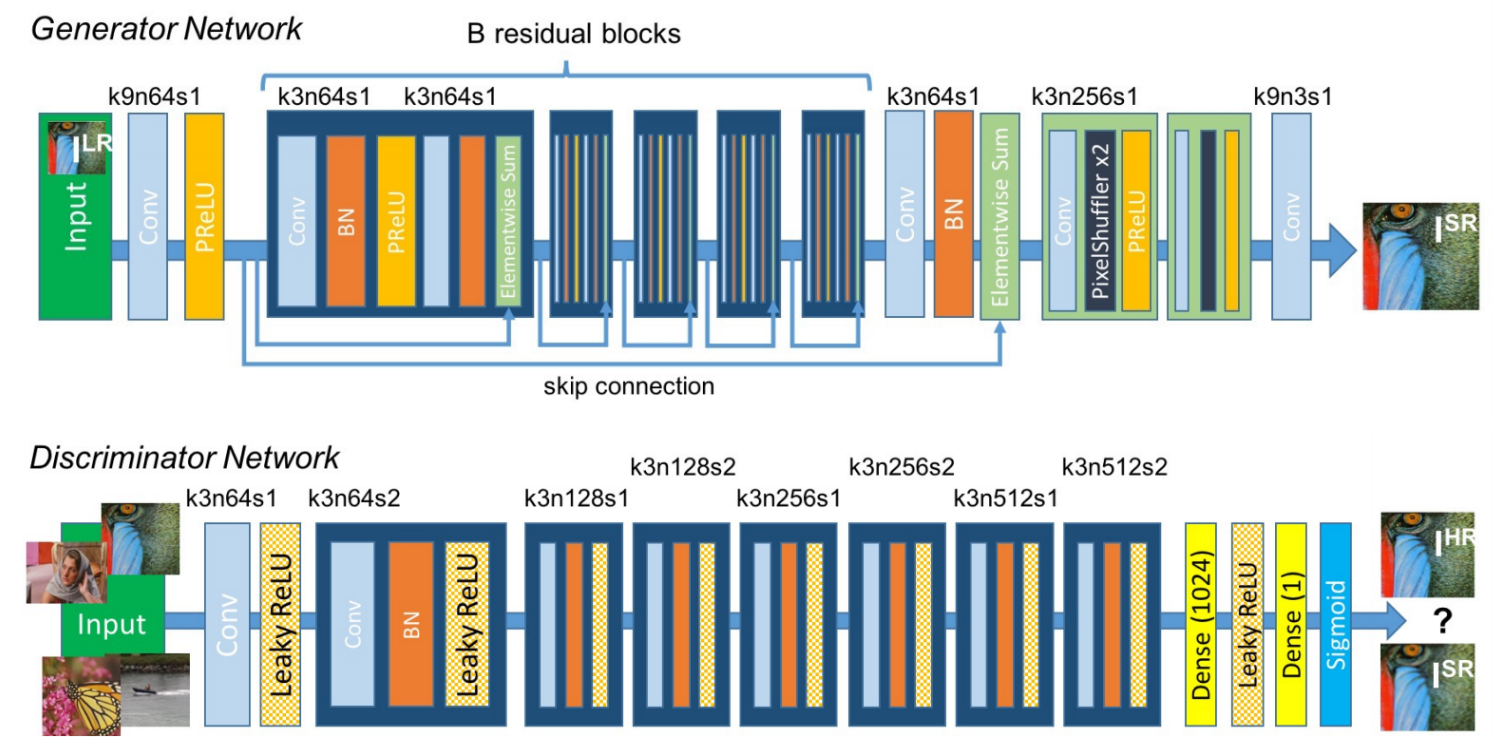} 
  \caption{The SRGAN model proposed in \cite{12}.}
  \label{fig:6}
\end{figure}

As its name suggests, the superresolution GAN (SRGAN) \cite{12} is a GAN model for performing image superresolution, as shown in Fig. \ref{fig:6}. The topology of the generator is a sequence of residual blocks, each of which is composed of a convolutional layer, a batch normalization layer, and a PReLU activation function, followed by another convolution, another batch normalization layer, and a summation layer, where residuals are added. In the original article, 16 residual blocks were used \cite{12}, but the number of residual blocks can vary according to the selected implementation. The SRGAN model was developed for image superresolution, so its last layers aim to increase the image size; however, the generator network is a fully convolutional network that it can address any input image size.

The discriminator network topology is also composed of residual blocks but uses the leaky ReLU activation function. It has dense layers at the end of its convolutional blocks, starting with 1024 neurons, and a sigmoid activation function at the end of the network, providing a way to differentiate between a true image obtained from the database and a synthetic image created by the generator.

The generator model of the SRGAN is trained with a loss function called the perceptual loss, which is computed using two terms, a content loss and an adversarial loss,

\begin{equation}
    L = L_{2}^{\phi} + 10^{-3} L_{G}.
\end{equation}

The generator loss $L_{G}$ is the same as that proposed in the original GAN paper \cite{1}. The content loss $L_{2}$ is the mean squared error (MSE) computed on the feature map. In the original article \cite{12}, the authors proposed to compute thecomputing $L_{2}$ based on the basis of the feature map generated by a Visual Geometry Group 19 (VGG19) \cite{20} neural network. The discriminator model is was trained to maximize the discriminator loss of the original GAN \cite{1}.

The SRGAN model can be adapted for denoising problems. In this case, the input of the generator is a noisy image, and the model is trained to generate a denoised image in such a way that the discriminator cannot identify whether the image is clean or has been cleaned by the model. The upsampling layers must be removed since the clean image in the output should have the same size as that of the noisy input image.

The adversarial training process of the complete SRGAN model is very complex and requires a significantly large amount of computational power. Preliminary tests showed that the denoising model could be deployed using only the generator, which had the same design as that of the SRGAN but was trained through standard supervised learning via the $L_{2}$ loss,

\begin{equation}
\label{eq:10}
    L_{2} = \dfrac{1}{Nm} \sum_{t=1}^{N} \sum_{t \in c_{i}} (y(t) - \hat{y(t)})^{2},
\end{equation}

where $y(t)$ is the target value, $\hat{y(t)}$ is the model prediction obtained for each value of crop $c_{i}$, $m=16,384$ is the number for a $256\times64$ crop, and $N$ is the total number of records included in the training dataset. This model is referred to as the SRGEN model.

The SRGEN topology with 4 residual blocks (SRGEN-4) was evaluated in this work. The details are shown in Table \ref{tab:table4}, with the same columns as those in Table \ref{tab:table3}, where arrows represent residual connections. The output layer was adapted for the denoising application via 64 $1\times1$ filters and a linear activation function, similar to the FCNN-5 model. The SRGEN-4 model contained a total of 1,098369 trainable parameters.

The SRGEN-4 model was trained using 4,000,000 records as training data and a batch size of 256 records. The optimizer was adaptive moment estimation (ADAM) with an initial learning rate of $10^{-5}$ and early stopping with a patience level of 50 epochs. The training procedure was halted if the maximum number of epochs (500) was reached or if the processing time exceeded 10 days. The ADAM optimizer was chosen as in the original paper \cite{12}, and it produced better results than those of SGD in preliminary tests.

\begin{table}[htpb]
\centering
\caption{SRGEN-4 Model Topology}
\label{tab:table4}
\begin{tabular}{lllp{60pt}}
\hline\noalign{\smallskip}
\textbf{Layer (Type)} & \textbf{Filter} & \textbf{Output Shape} & \textbf{Parameters} \\
\noalign{\smallskip}\hline\noalign{\smallskip}
Conv2d-1    & $9\times9$    & {[}64, 256, 64{]}   & 5,248 \\
PReLU-2     & -             & {[}64, 256, 64{]}   & 64    \\
\multicolumn{4}{l}{\textbf{RESIDUAL BLOCK 1}} \\ \noalign{\smallskip}
Conv2d-3    & $3\times3$    & {[}64, 256, 64{]}   & 36,928 \\
BatchNorm2d-4 & -           & {[}64, 256, 64{]}   & 128    \\
PReLU-5     & -             & {[}64, 256, 64{]}   & 64     \\
Conv2d-6    & $3\times3$    & {[}64, 256, 64{]}   & 36,928 \\
BatchNorm2d-7 & -           & {[}64, 256, 64{]}   & 128    \\
PReLU-8     & -             & {[}64, 256, 64{]}   & 64     \\
Residual sum & -            & {[}64, 256, 64{]}   & -      \\
\multicolumn{4}{l}{\textbf{RESIDUAL BLOCK 2}} \\ \noalign{\smallskip}
Conv2d-9    & $3\times3$    & {[}64, 256, 64{]}   & 36,928 \\
BatchNorm2d-10 & -          & {[}64, 256, 64{]}   & 128    \\
PReLU-11    & -             & {[}64, 256, 64{]}   & 64     \\
Conv2d-12   & $3\times3$    & {[}64, 256, 64{]}   & 36,928 \\
BatchNorm2d-13 & -          & {[}64, 256, 64{]}   & 128    \\
PReLU-14    & -             & {[}64, 256, 64{]}   & 64     \\
Residual sum & -            & {[}64, 256, 64{]}   & -      \\
\multicolumn{4}{l}{\textbf{RESIDUAL BLOCK 3}} \\ \noalign{\smallskip}
Conv2d-15   & $3\times3$    & {[}64, 256, 64{]}   & 36,928 \\
BatchNorm2d-16 & -          & {[}64, 256, 64{]}   & 128    \\
PReLU-17    & -             & {[}64, 256, 64{]}   & 64     \\
Conv2d-18   & $3\times3$    & {[}64, 256, 64{]}   & 36,928 \\
BatchNorm2d-19 & -          & {[}64, 256, 64{]}   & 128    \\
PReLU-20    & -             & {[}64, 256, 64{]}   & 64     \\
Residual sum & -            & {[}64, 256, 64{]}   & -      \\
\multicolumn{4}{l}{\textbf{RESIDUAL BLOCK 4}} \\ \noalign{\smallskip}
Conv2d-21   & $3\times3$    & {[}64, 256, 64{]}   & 36,928 \\
BatchNorm2d-22 & -          & {[}64, 256, 64{]}   & 128    \\
PReLU-23    & -             & {[}64, 256, 64{]}   & 64     \\
Conv2d-24   & $3\times3$    & {[}64, 256, 64{]}   & 36,928 \\
BatchNorm2d-25 & -          & {[}64, 256, 64{]}   & 128    \\
PReLU-26    & -             & {[}64, 256, 64{]}   & 64     \\
Residual sum & -            & {[}64, 256, 64{]}   & -      \\
Conv2d-27   & $3\times3$    & {[}64, 256, 64{]}   & 36,928 \\
BatchNorm2d-28 & -          & {[}64, 256, 64{]}   & 128    \\
Residual sum & -            & {[}64, 256, 64{]}   & -      \\
Conv2d-29   & $3\times3$    & {[}256, 256, 64{]}  & 147,712 \\
PReLU-30    & -             & {[}256, 256, 64{]}  & 256    \\
Conv2d-31   & $9\times9$    & {[}256, 256, 64{]}  & 590,080 \\
PReLU-32    & -             & {[}256, 256, 64{]}  & 256    \\
Conv2d-33   & $1\times1$    & {[}1, 256, 64{]}    & 20,737 \\
\noalign{\smallskip}\hline
\end{tabular}
\end{table}

\section{Results}
\label{sec:5}
The generalization capacities of the models were evaluated with different combinations of training and testing sets so that different structural models and noise files were used for training and testing. For example, if one of the MARMOUSI (SEAM) structural models was used for training, then one of the SEAM (MARMOUSI) structural models was used for testing, and if the NOISE 1 (NOISE 2) file was used for training, then the NOISE 2 (NOISE 1) file was used for testing. Moreover, for the same structural model-noise file combination, different levels of noise were used for training and testing. For example, if L1 was used for testing, then the model was trained with levels L2, L5, and L10 such that the same noise level was never present in the training and testing data. All combinations of the training and testing sets resulted in 32 experiments, as shown in Table \ref{tab:tableA1} in the appendix, which were executed for each DL model.

The FCNN-5 and SRGEN-4 models evaluated in this work were implemented in Python 3 via the PyTorch framework. All the experiments described in this section were carried out on the Petrobras Gaia supercomputer, in which each computing node had 1 TB of main RAM and 4 NVIDIA A100 GPUs with 80 GB of memory each. All the experiments were terminated after the threshold of 10 days of processing time was reached.

\subsection{Evaluation metrics}

The peak SNR (PSNR) is a commonly used metric for image denoising evaluations. The PSNR is computed as

\begin{equation}
    \textnormal{PSNR} = 10 log_{10} \dfrac{\textnormal{R}^{2}}{\textnormal{MSE}},
\end{equation}

where $\textnormal{R}$ is a scale factor that represents the maximum value that a signal can achieve in the utilized representation, and the $\textnormal{MSE}$ is the mean squared error, which is computed as in Eq. \ref{eq:10}.

In the literature, models are usually evaluated on 8-bit images such that $\textnormal{R}=256$ . In more general settings, $\textnormal{R}=2^{n}$ , where $n$ is the number of bits in the representation. In a case involving seismic data, the shot gathers are represented by real numbers, usually floating points, so the maximum value in the representation does not make any sense for the PSNR calculation. In this work, a fixed scale factor of  was used to compare several different files, considering that the data were scaled according to the modified z score method (cf. Eq. \ref{eq:7}).

The PSNR is not very useful for comparing DL models, as the differences between the models are small and are not reflected in the values of this metric. A relative measure was adopted for the comparison, that is, the SNR relative ratio ($\textnormal{SNR}^{2}$), and it was computed as the ratio of the SNRs of the original (noisy) signal and the signal predicted by the model as

\begin{equation}
    \textnormal{SNR}^{2} = 1 - \dfrac{\textnormal{SNR}(\hat{y})}{\textnormal{SNR}(y)} = 1 - \dfrac{\textnormal{RMS}(e_{M})}{\textnormal{RMS}(e_{R})},
\end{equation}

where $e_{M}$ is the model residue and $e_{R}$ is the noise contained in the noisy data.

The $\textnormal{SNR}^{2}$ is interpreted similarly to the coefficient of determination $\textnormal{R}^{2}$ as a relation between the model error and the variance of the residuals. When the $\textnormal{SNR}^{2}$ value is smaller than 1.0, the higher the value is, the better. It is possible for $\textnormal{SNR}^{2}$ values to be negative, and this is the case when, for example, the energy removed from the signal is greater than the energy of the original noise.

\subsection{Model results}

The evaluation metrics computed for all 32 combinations of training and testing sets are shown in Table 5 for the FCNN-5 model and in Table 6 for the SRGEN-4 model. The results shown in Table 5 and Table 6 correspond to the averages and standard deviations, respectively, for all shot gathers in the testing set. The columns referred to as L1 - L10 in Table \ref{tab:table5} and Table \ref{tab:table6} indicate the noise levels used in the testing set, which were not used in the training set (cf. Table \ref{tab:tableA1}).

\begin{figure}[ht]
  \centering
    \includegraphics[width=1.0\textwidth]{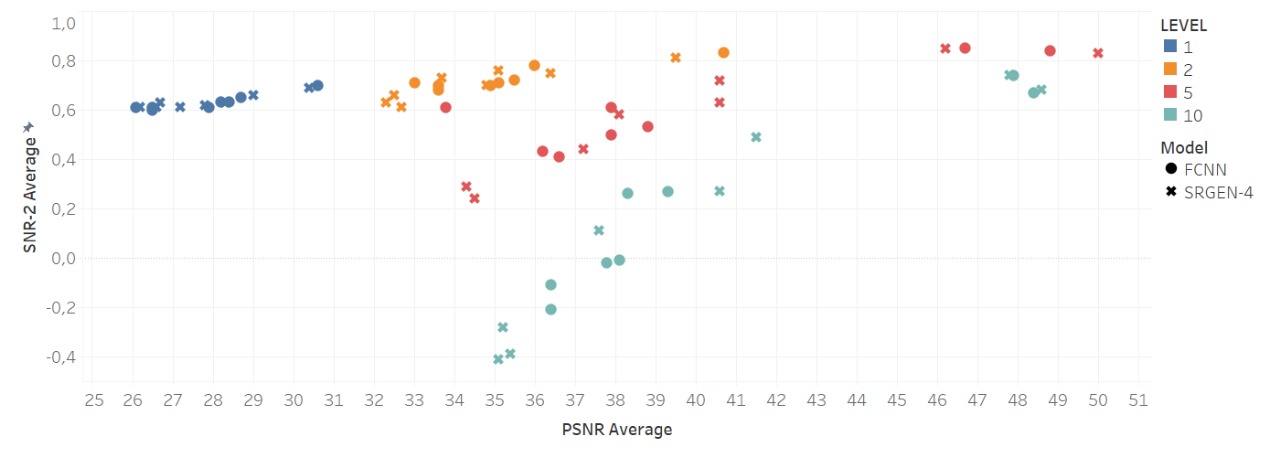} 
  \caption{Comparison among the attained evaluation metrics.}
  \label{fig:7}
\end{figure}

A comparison among the evaluation metrics is shown in Fig. \ref{fig:7} for the average $\textnormal{PSNR}$ and $\textnormal{SNR}^{2}$ values across all the shot gathers. The results in Fig. \ref{fig:7} show that the $\textnormal{PSNR}$ is not very suitable as an evaluation metric, as most of the experiments fell within the same range of 32 dB to 42 dB, except when the L1 noise level was utilized for the testing set, which produced lower PSNR results, and some experiments (mostly those involving the SEISMIC B structural model) using the L5 and L10 noise levels in the testing set, which yielded higher $\textnormal{PSNR}$ values. The $\textnormal{SNR}^{2}$ evaluation metric is used to present and discuss the model results.

\begin{landscape}

\begin{table}[ht!]
\centering
\caption{FCNN-5 model results}
\label{tab:table5}
\resizebox{1.2\textwidth}{!}{%
\begin{tabular}{lll lll llll llll}
\hline\noalign{\smallskip}
\multicolumn{2}{l}{\textbf{TRAIN}} & \multicolumn{2}{l}{\textbf{TEST}} & \multicolumn{4}{l}{\textbf{PSNR}} & \multicolumn{4}{l}{\textbf{SNR$^{2}$}} \\
\noalign{\smallskip}\hline\noalign{\smallskip}
\textbf{SEISMIC} & \textbf{NOISE} & \textbf{SEISMIC} & \textbf{NOISE} & \textbf{L1} & \textbf{L2} & \textbf{L5} & \textbf{L10} & \textbf{L1} & \textbf{L2} & \textbf{L5} & \textbf{L10} \\
\noalign{\smallskip}\hline\noalign{\smallskip}
SEISMIC A & NOISE 1 & SEISMIC B & NOISE 2 & $30.6 \pm 3.2$ & $40.7 \pm 4.2$ & $48.8 \pm 3.0$ & $48.4 \pm 1.9$ & $0.70 \pm 0.07$ & $0.83 \pm 0.06$ & $0.84 \pm 0.04$ & $0.67 \pm 0.07$ \\
SEISMIC A & NOISE 2 & SEISMIC B & NOISE 1 & $26.5 \pm 2.6$ & $36.0 \pm 3.8$ & $46.7 \pm 2.7$ & $47.9 \pm 2.0$ & $0.61 \pm 0.05$ & $0.78 \pm 0.05$ & $0.85 \pm 0.03$ & $0.74 \pm 0.07$ \\
SEISMIC B & NOISE 1 & SEISMIC A & NOISE 2 & $28.2 \pm 2.3$ & $35.5 \pm 1.9$ & $36.6 \pm 0.5$ & $36.4 \pm 0.2$ & $0.63 \pm 0.04$ & $0.72 \pm 0.02$ & $0.41 \pm 0.11$ & $-0.21 \pm 0.26$ \\
SEISMIC B & NOISE 2 & SEISMIC A & NOISE 1 & $27.9 \pm 3.0$ & $33.6 \pm 2.1$ & $36.2 \pm 0.5$ & $36.4 \pm 0.2$ & $0.61 \pm 0.05$ & $0.68 \pm 0.02$ & $0.43 \pm 0.13$ & $-0.11 \pm 0.30$ \\
SEISMIC C & NOISE 1 & SEISMIC D & NOISE 2 & $28.7 \pm 3.1$ & $35.1 \pm 2.8$ & $37.9 \pm 1.7$ & $37.8 \pm 1.4$ & $0.65 \pm 0.06$ & $0.71 \pm 0.04$ & $0.50 \pm 0.09$ & $-0.02 \pm 0.21$ \\
SEISMIC C & NOISE 2 & SEISMIC D & NOISE 1 & $26.1 \pm 2.4$ & $33.0 \pm 2.7$ & $37.9 \pm 1.5$ & $38.3 \pm 1.3$ & $0.61 \pm 0.05$ & $0.71 \pm 0.04$ & $0.61 \pm 0.06$ & $0.26 \pm 0.14$ \\
SEISMIC D & NOISE 1 & SEISMIC C & NOISE 2 & $28.4 \pm 2.3$ & $34.9 \pm 2.1$ & $38.8 \pm 0.9$ & $38.1 \pm 0.4$ & $0.63 \pm 0.04$ & $0.70 \pm 0.02$ & $0.53 \pm 0.08$ & $-0.01 \pm 0.20$ \\
SEISMIC D & NOISE 2 & SEISMIC C & NOISE 1 & $26.5 \pm 2.5$ & $33.6 \pm 2.4$ & $33.8 \pm 0.8$ & $39.3 \pm 0.3$ & $0.60 \pm 0.05$ & $0.70 \pm 0.02$ & $0.61 \pm 0.07$ & $0.27 \pm 0.28$ \\
\noalign{\smallskip}\hline
\end{tabular}%
}
\end{table}


\begin{table}[ht!]
\centering
\caption{SRGEN-4 model results}
\label{tab:table6}
\resizebox{1.2\textwidth}{!}{%
\begin{tabular}{llllllllllllll}
\hline\noalign{\smallskip}
\textbf{TRAIN} & \textbf{TEST} & \textbf{SEISMIC} & \textbf{NOISE} & \textbf{L1} & \textbf{L2} & \textbf{L5} & \textbf{L10} & \textbf{L1} & \textbf{L2} & \textbf{L5} & \textbf{L10} \\
\noalign{\smallskip}\hline\noalign{\smallskip}
SEISMIC A & NOISE 1 & SEISMIC B & NOISE 2 & $30.4 \pm 2.9$ & $39.5 \pm 3.5$ & $50.0 \pm 2.3$ & $48.6 \pm 1.4$ & $0.69 \pm 0.06$ & $0.81 \pm 0.04$ & $0.83 \pm 0.03$ & $0.68 \pm 0.06$ \\
SEISMIC A & NOISE 2 & SEISMIC B & NOISE 1 & $26.7 \pm 2.7$ & $35.1 \pm 3.3$ & $46.2 \pm 2.1$ & $47.8 \pm 0.5$ & $0.63 \pm 0.05$ & $0.76 \pm 0.04$ & $0.85 \pm 0.01$ & $0.74 \pm 0.06$ \\
SEISMIC B & NOISE 1 & SEISMIC A & NOISE 2 & $27.8 \pm 2.1$ & $32.7 \pm 1.4$ & $34.5 \pm 0.4$ & $35.1 \pm 0.2$ & $0.62 \pm 0.03$ & $0.61 \pm 0.03$ & $0.24 \pm 0.14$ & $-0.41 \pm 0.29$ \\
SEISMIC B & NOISE 2 & SEISMIC A & NOISE 1 & $27.2 \pm 2.5$ & $32.3 \pm 1.5$ & $34.3 \pm 0.2$ & $35.2 \pm 0.2$ & $0.61 \pm 0.04$ & $0.63 \pm 0.04$ & $0.29 \pm 0.19$ & $-0.28 \pm 0.36$ \\
SEISMIC C & NOISE 1 & SEISMIC D & NOISE 2 & $29.0 \pm 3.2$ & $36.4 \pm 3.3$ & $40.6 \pm 2.0$ & $40.6 \pm 1.5$ & $0.66 \pm 0.06$ & $0.75 \pm 0.04$ & $0.63 \pm 0.06$ & $0.27 \pm 0.15$ \\
SEISMIC C & NOISE 2 & SEISMIC D & NOISE 1 & $26.2 \pm 2.5$ & $33.7 \pm 2.9$ & $40.6 \pm 1.8$ & $41.5 \pm 1.3$ & $0.61 \pm 0.05$ & $0.73 \pm 0.04$ & $0.72 \pm 0.04$ & $0.49 \pm 0.09$ \\
SEISMIC D & NOISE 1 & SEISMIC C & NOISE 2 & $28.3 \pm 2.2$ & $34.8 \pm 2.0$ & $37.2 \pm 0.9$ & $35.4 \pm 0.5$ & $0.63 \pm 0.04$ & $0.70 \pm 0.02$ & $0.44 \pm 0.10$ & $-0.39 \pm 0.30$ \\
SEISMIC D & NOISE 2 & SEISMIC C & NOISE 1 & $26.6 \pm 2.5$ & $32.5 \pm 2.0$ & $38.1 \pm 0.9$ & $37.6 \pm 0.5$ & $0.61 \pm 0.05$ & $0.66 \pm 0.02$ & $0.58 \pm 0.08$ & $0.11 \pm 0.21$ \\
\noalign{\smallskip}\hline
\end{tabular}%
}
\end{table}

\end{landscape}


\begin{figure}[ht]
  \centering
  \includegraphics[width=1.0\textwidth]{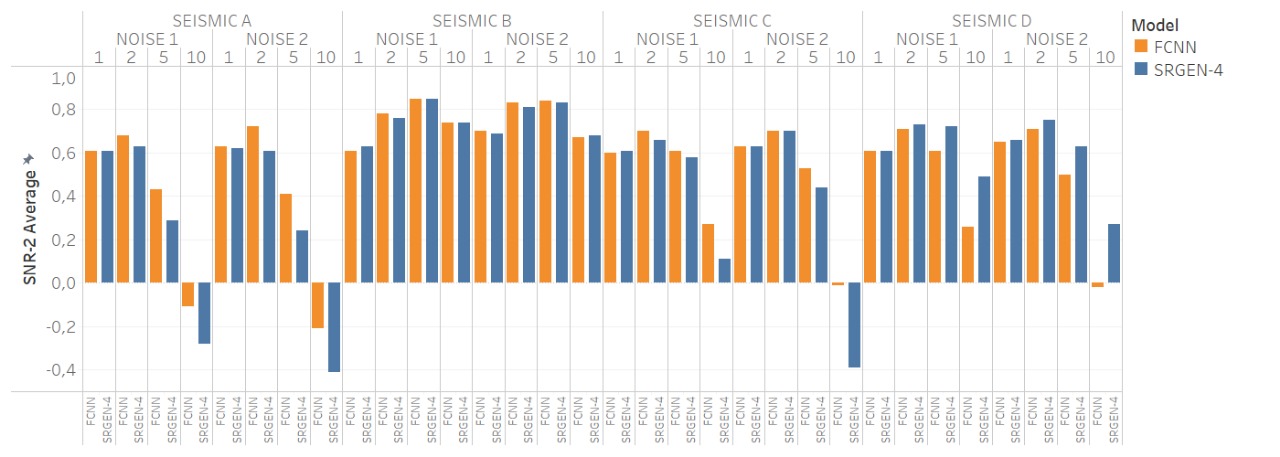} 
  \caption{Comparison among model results based on the average $\textnormal{SNR}^{2}$ values.}
  \label{fig:8}
\end{figure}

\begin{figure}[ht]
  \centering
  \includegraphics[width=1.0\textwidth]{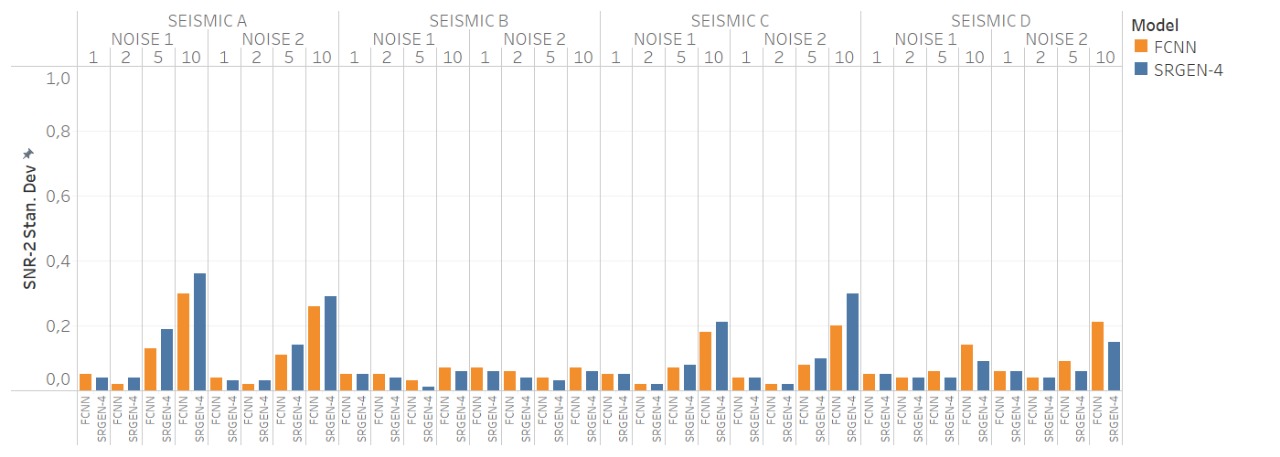} 
  \caption{Comparison among the model results based on the standard deviation of the $\textnormal{SNR}^{2}$ values.}
  \label{fig:9}
\end{figure}

The results in Table \ref{tab:table5} and Table \ref{tab:table6} are shown graphically in Figs. \ref{fig:8} and \ref{fig:9}, respectively where Fig. \ref{fig:8} shows the average $\textnormal{SNR}^{2}$ among all the shot gathers and Fig. \ref{fig:9} shows the standard deviation of the $\textnormal{SNR}^{2}$ values among all the shot gathers.

Figures \ref{fig:8} and \ref{fig:9} show that FCNN-5 and SRGEN-4 achieved similar performances in all the experiments, with the worst results obtained for the higher noise levels by both models. The SRGEN-4 model has a more complex topology with some improvements, such as batch normalization and residual connections, but it has almost the same number of trainable parameters as the FCNN-5 model does. However, a qualitative evaluation of the seismograms obtained for these models revealed some visual model performance differences that were not captured by the metrics, as discussed in the next section.

\subsection{Discussion}

\begin{figure}[ht!]
    \centering
    \begin{subfigure}[b]{0.4\linewidth}
        \centering
        \includegraphics[width=\linewidth]{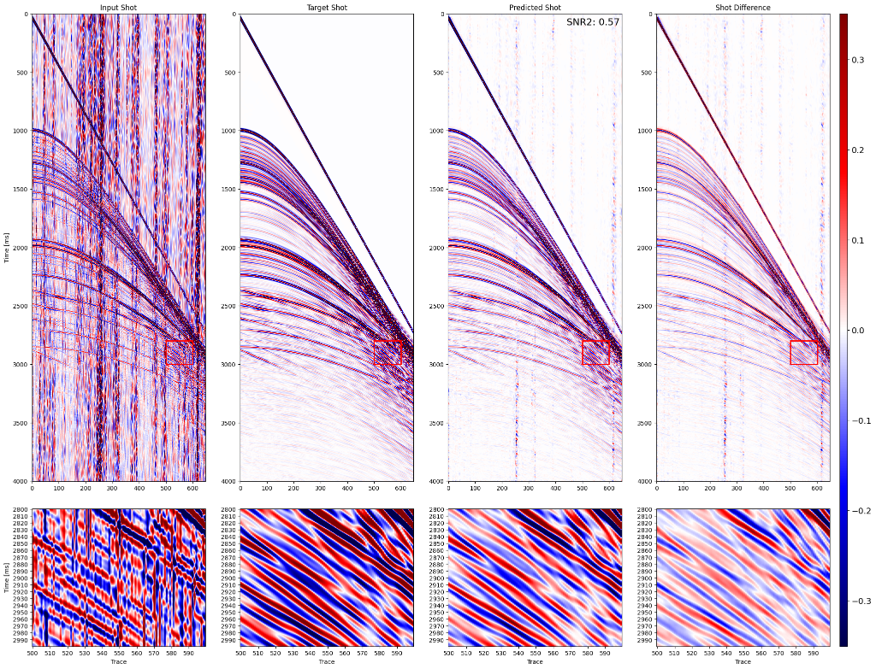}
        \caption{}
    \end{subfigure}
    \begin{subfigure}[b]{0.4\linewidth}
        \centering
        \includegraphics[width=\linewidth]{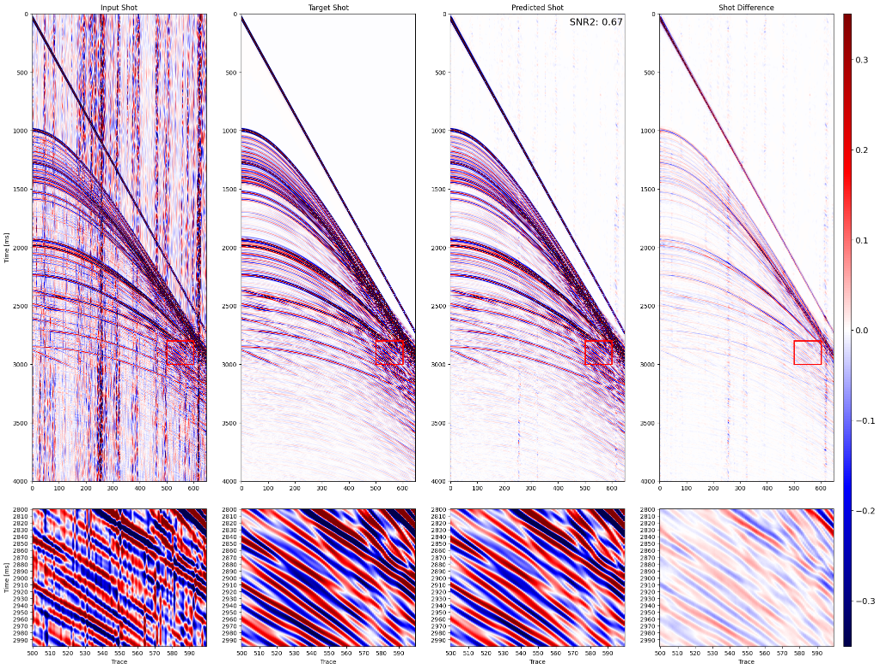}
        \caption{}
    \end{subfigure}
    \\
    \begin{subfigure}[b]{0.4\linewidth}
        \centering
        \includegraphics[width=\linewidth]{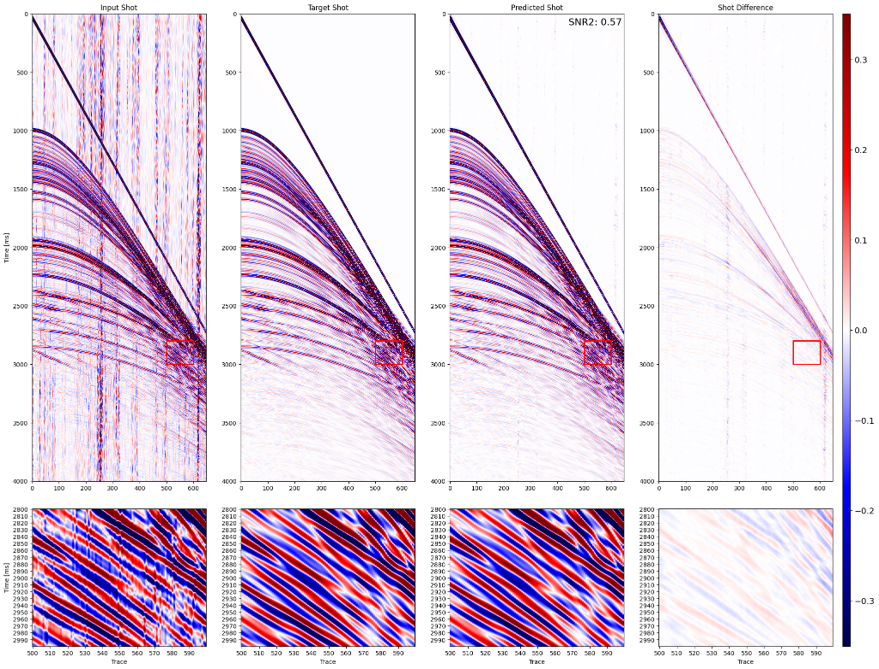}
        \caption{}
    \end{subfigure}
    \begin{subfigure}[b]{0.4\linewidth}
        \centering
        \includegraphics[width=\linewidth]{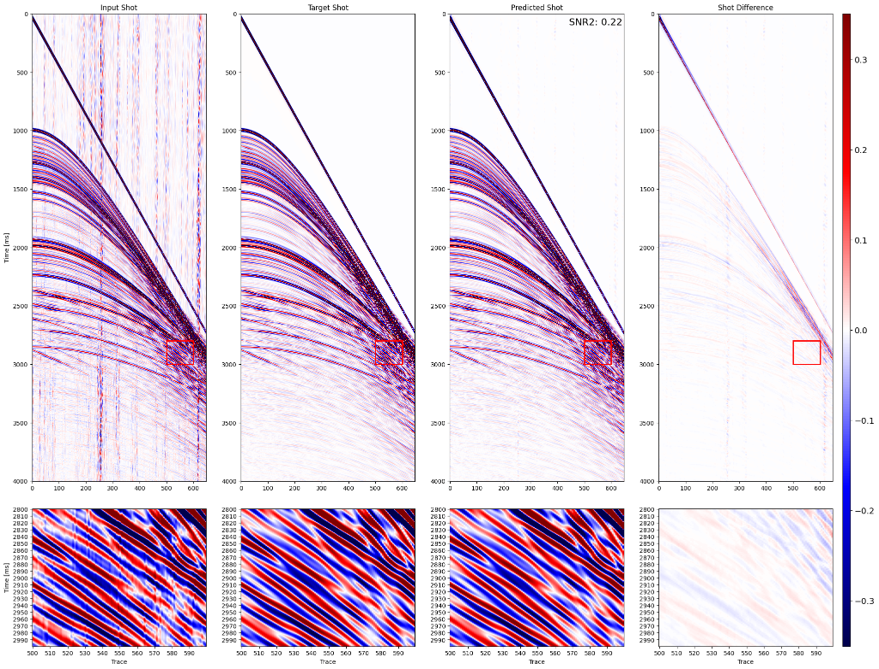}
        \caption{}
    \end{subfigure}
    \caption{FCNN-5 results for SEISMIC A and NOISE 1: (a) L1, (b) L2, (c) L5, (d) L10.}
    \label{fig:10}
\end{figure}

Figure \ref{fig:10} shows the seismograms of one shot gather of the SEISMIC A synthetic (clean) data combined with the NOISE 1 file at the four noise levels and the FCNN-5 model results. Each of the subfigures shows four seismograms from left to right: the noisy data, clean data, model predictions and residuals, in other words, the difference between the clean data and the model predictions. Moreover, in the second line of each subfigure, the region represented by the red square is zoomed in. The color palette in Fig. \ref{fig:10} is the same for all the seismograms and is defined as the range between percentiles P2 and P98 of the clean data. A file with similar visualizations for all the experiments and the results of the two models is available in the supplementary material in the data repository (see data availability section).

The FCNN-5 model was successful at attenuating noise when the noise signal was stronger (L1 and L2) and almost completely eliminated noise when the noise signal was weaker (L5 and L10). The zoomed-in version of each subfigure shows that the signal had good quality after denoising, even for stronger noise levels. However, the model also removed signals when conducting denoising, and the stronger the noise level was, the more signal was removed.

\begin{figure}[ht!]
    \centering
    \begin{subfigure}[b]{0.4\linewidth}
        \centering
        \includegraphics[width=\linewidth]{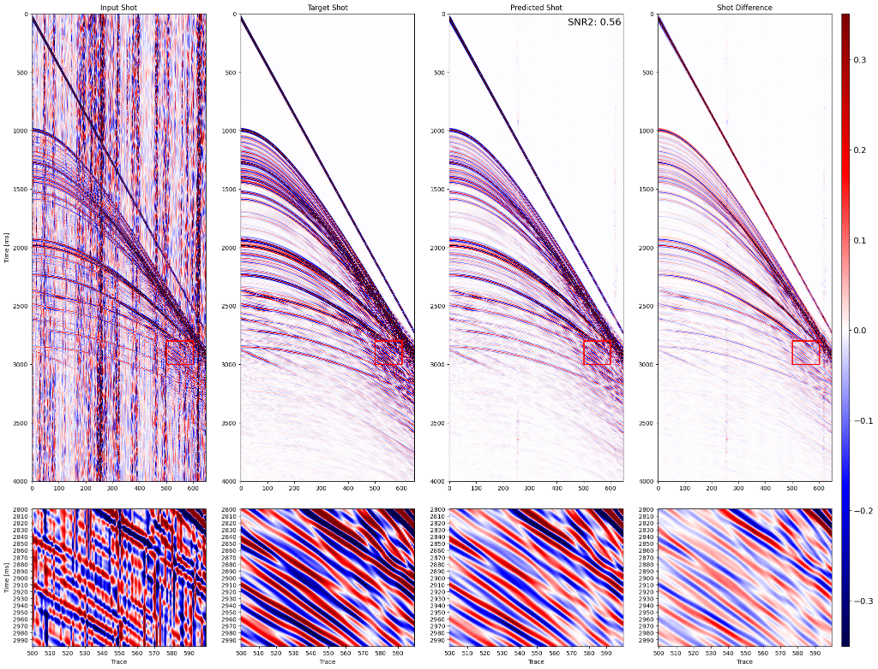}
        \caption{}
    \end{subfigure}
    \begin{subfigure}[b]{0.4\linewidth}
        \centering
        \includegraphics[width=\linewidth]{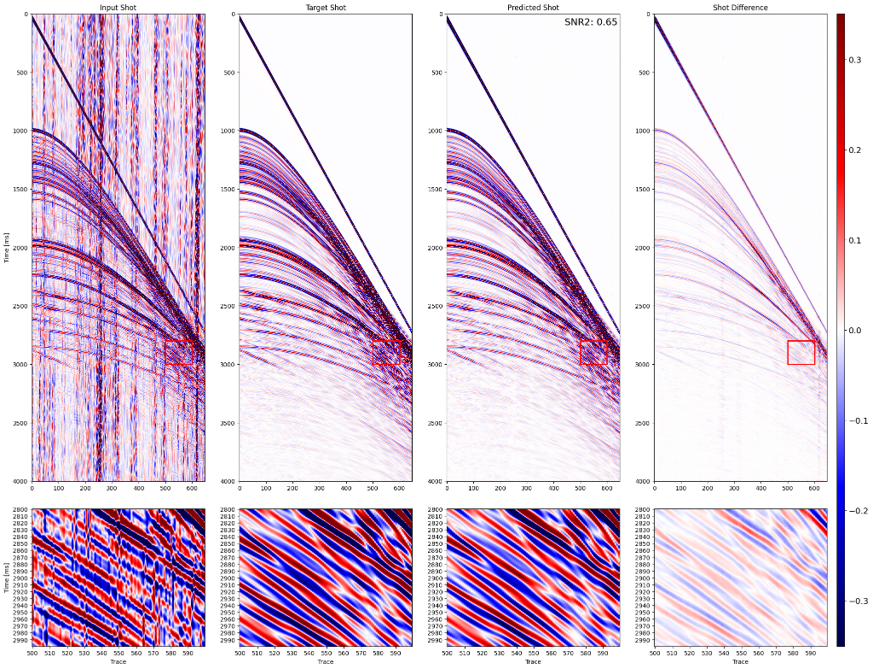}
        \caption{}
    \end{subfigure}
    \\
    \begin{subfigure}[b]{0.4\linewidth}
        \centering
        \includegraphics[width=\linewidth]{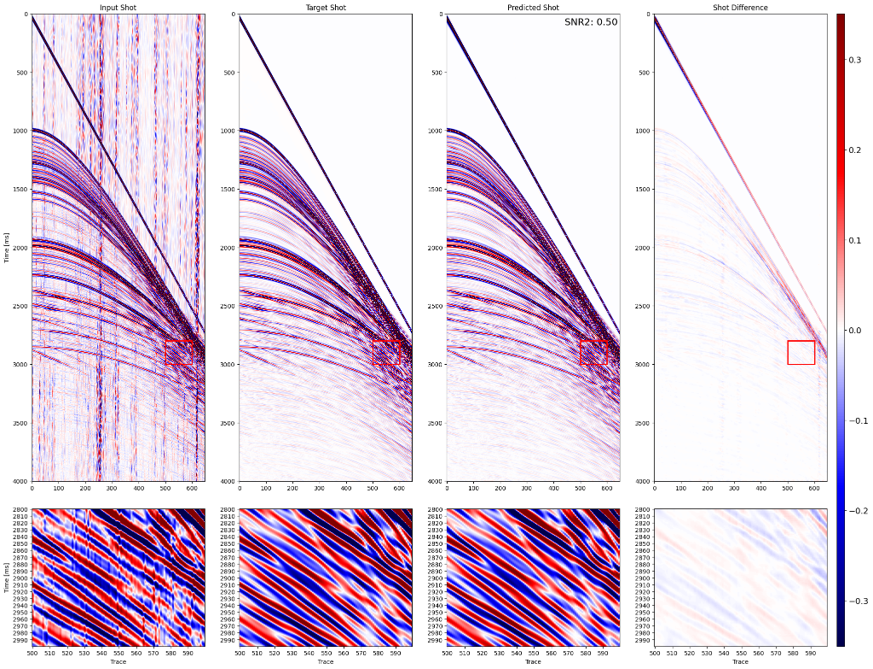}
        \caption{}
    \end{subfigure}
    \begin{subfigure}[b]{0.4\linewidth}
        \centering
        \includegraphics[width=\linewidth]{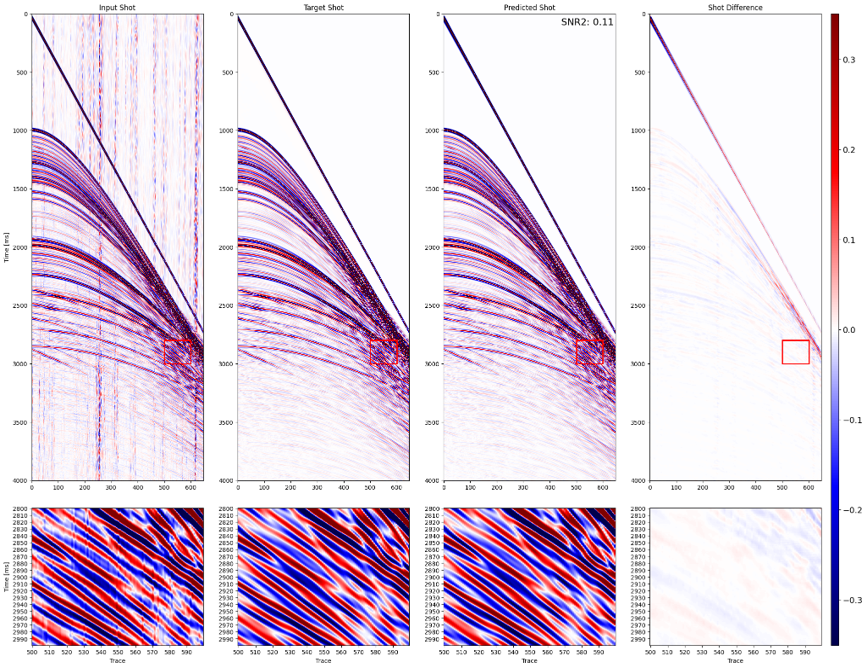}
        \caption{}
    \end{subfigure}
    \caption{SRGEN-4 results for SEISMIC A and NOISE 1: (a) L1, (b) L2, (c) L5, (d) L10.}
    \label{fig:11}
\end{figure}

Figure \ref{fig:11} shows the results produced by the SRGEN-4 model for the same shot gather shown in Fig. \ref{fig:10}, and the subfigures are organized in the same order as those in Fig. \ref{fig:10}. Visually, the SRGEN-4 model was more successful at removing noise even in cases with stronger noise levels. Some signals were also removed when denoising.

The $\textnormal{SNR}^{2}$ values computed for the shot gathers presented in Fig. \ref{fig:10} and Fig. \ref{fig:11} are shown in Table \ref{tab:table7}. Although the results seem to be visually different, with slightly better performance for the SRGEN-4 model, the $\textnormal{SNR}^{2}$ results are very similar, except for those of the experiments with the L10 noise level in the testing set, where the SRGEN-4 model performs better.

\begin{table}[htpb]
\centering
\caption{$\textnormal{SNR}^{2}$ results obtained for the seismograms presented in Figure \ref{fig:10} and Figure \ref{fig:11}}
\label{tab:table7}
\begin{tabular}{lcccc}
\hline\noalign{\smallskip}
         & \textbf{L1} & \textbf{L2} & \textbf{L5} & \textbf{L10} \\
\noalign{\smallskip}\hline\noalign{\smallskip}
\textbf{FCNN-5}  & 0.57        & 0.67        & 0.57        & 0.22         \\ 
\textbf{SRGEN-4} & 0.56        & 0.65        & 0.50        & 0.11         \\
\noalign{\smallskip}\hline
\end{tabular}
\end{table}

These results seem to be intriguing, as the metrics do not correspond to the visual evaluation outcomes of the figure. In fact, the results show that $\textnormal{SNR}^{2}$ is very sensitive to the amount of signal that is removed by the denoising process. When the energy of the removed signal is important or at the same order of the energy of the noise, the $\textnormal{SNR}^{2}$ value becomes very low, even if the seismogram shows an almost clean image. Moreover, some reflections, such as direct waves, have very high amplitudes that were not effectively reconstructed by the model and dominated the error. Thus, to obtain a higher $\textnormal{SNR}^{2}$ value, the model must remove the noise without removing the signal. This is a challenging problem for DL model applications in seismic data denoising.

\section{Conclusion}
\label{sec:6}

This work introduces a seismic denoising benchmark dataset consisting of synthetic seismic data corrupted with real noise files extracted after executing a filtering process. The dataset is proposed as a benchmark for accelerating the development of new DL model solutions for seismic data denoising. The proposed benchmark dataset is used in this work to compare two well-known DL architectures. This work also introduces a new evaluation metric that can capture small variations in model results.

Seismic responses simulated from two different open-source structural models were used to generate four synthetic datasets that were used as clean seismic data. The synthetic datasets were combined with two real swell noise files at four different noise levels. The model evaluation were performed in 32 different experiments, such that different structural model-noise file combinations were used for training and testing, such that the same noise level was never present in the training and testing data of the same experiment. These 32 experiments can be used as benchmarks for new model evaluations, or the real noise files can be combined with other seismic data to generate new noisy seismic data.

Two DL models were used for denoising in this work, one fully convolutional neural network (FCNN-5) and the SRGEN-4 model, which was derived from the SRGAN model; both models were trained via supervised learning with L2 loss. The models were shown to be able to perform model denoising, even at high noise levels. Both models achieved similar performances in all the experiments when evaluated by the SNR2 metric. The SRGEN-4 model had a more complex topology with some improvements, such as batch normalization and residual connections, although it possessed roughly the same number of trainable parameters as that of the FCNN-5 model.

The results showed that $\textnormal{SNR}^{2}$ is very sensitive to the amount of signal that is removed by the denoising process; to obtain a higher $\textnormal{SNR}^{2}$ value, the model must remove the noise without removing the signal. This is a challenging problem for the DL model application in seismic data denoising, and the benchmark dataset presented in this work can be used to develop and compare new DL models for seismic data denoising.

\section{Data availability}
\label{sec:7}

The dataset is open and available at the Harvard Dataverse repository: https://doi.org/10.7910/DVN/EXOCYD. The data have been released under a Creative Commons (CC) license (CC BY-NC-SA), which enables users to distribute, remix, adapt, and build upon the material in any medium or format for noncommercial purposes only, and only so long as attribution is given to the creator. If another dataset is remixed, adapted, or built upon this dataset, the modified material must be licensed under identical terms.

\textbf{Acknowledgements:} The authors are grateful to Petrobras for their financial support, for providing the real swell noise data used in this work, and for providing the high-performance computing environment used to run the experiments. The authors are also grateful to the Brazilian Research Council – CNPq and the Rio de Janeiro State Research Agency – FAPERJ for their financial support for this research.

\bibliographystyle{unsrt}  
\bibliography{references}  






\appendix
\section*{Appendix}
\addcontentsline{toc}{section}{Appendix} 
\renewcommand{\thetable}{A.\arabic{table}}
\renewcommand{\thefigure}{A.\arabic{figure}}
\setcounter{table}{0}
\setcounter{figure}{0}

\begin{figure}[ht]
    \centering
    \begin{subfigure}[b]{0.24\textwidth}
        \includegraphics[width=\textwidth]{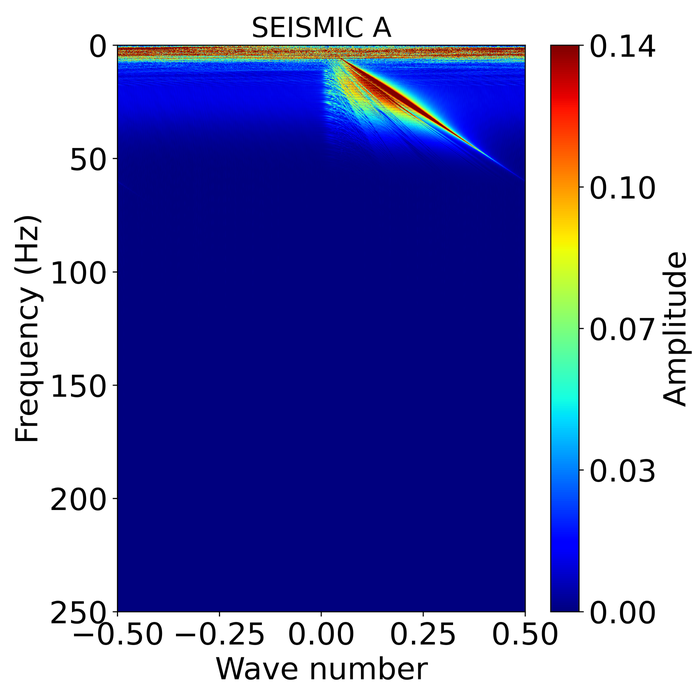}
    \end{subfigure}
    \hfill
    \begin{subfigure}[b]{0.24\textwidth}
        \includegraphics[width=\textwidth]{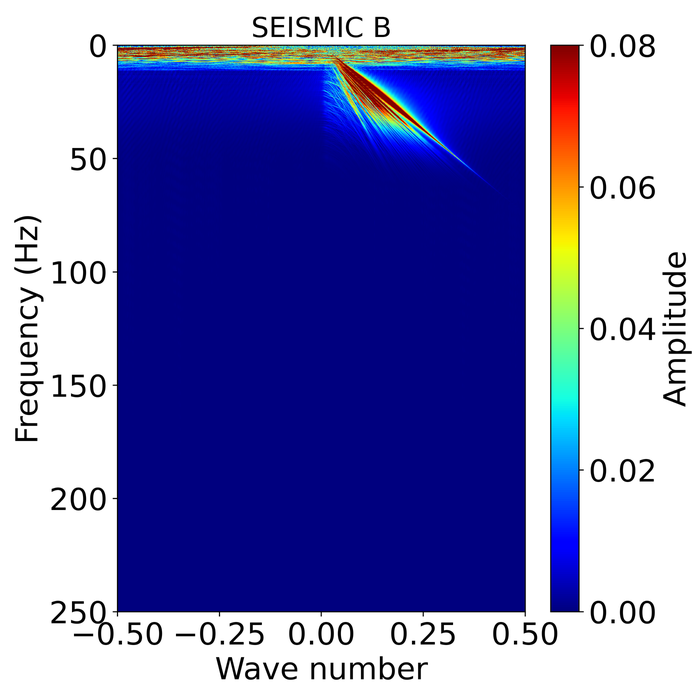}
    \end{subfigure}
    \hfill
    \begin{subfigure}[b]{0.24\textwidth}
        \includegraphics[width=\textwidth]{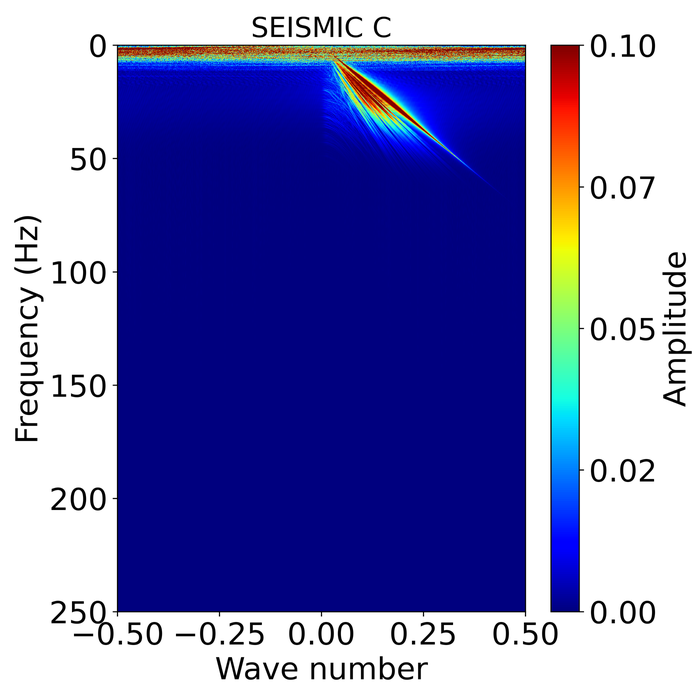}
    \end{subfigure}
    \hfill
    \begin{subfigure}[b]{0.24\textwidth}
        \includegraphics[width=\textwidth]{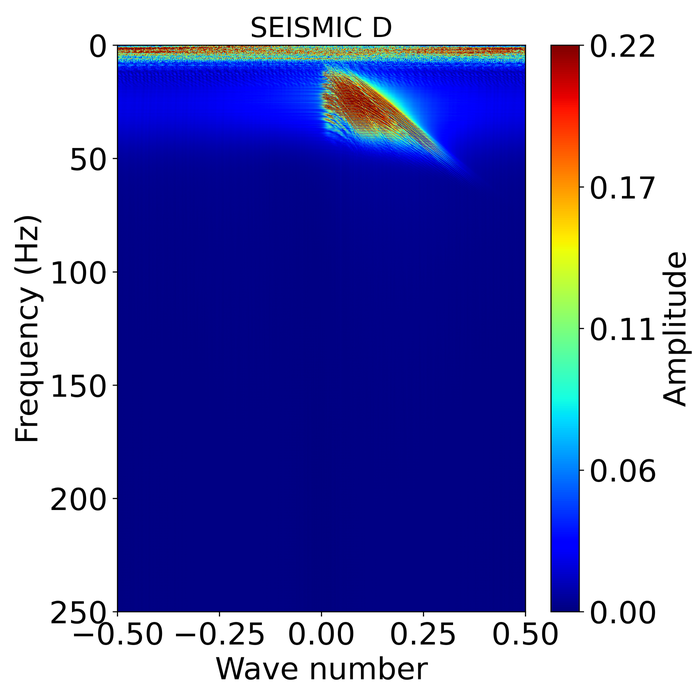}
    \end{subfigure}
    
    \vskip\baselineskip
    
    \begin{subfigure}[b]{0.24\textwidth}
        \includegraphics[width=\textwidth]{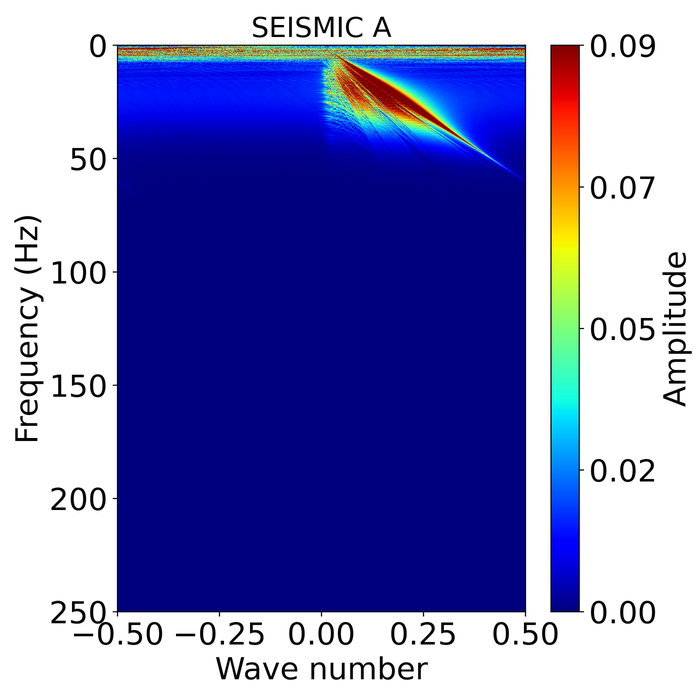}
    \end{subfigure}
    \hfill
    \begin{subfigure}[b]{0.24\textwidth}
        \includegraphics[width=\textwidth]{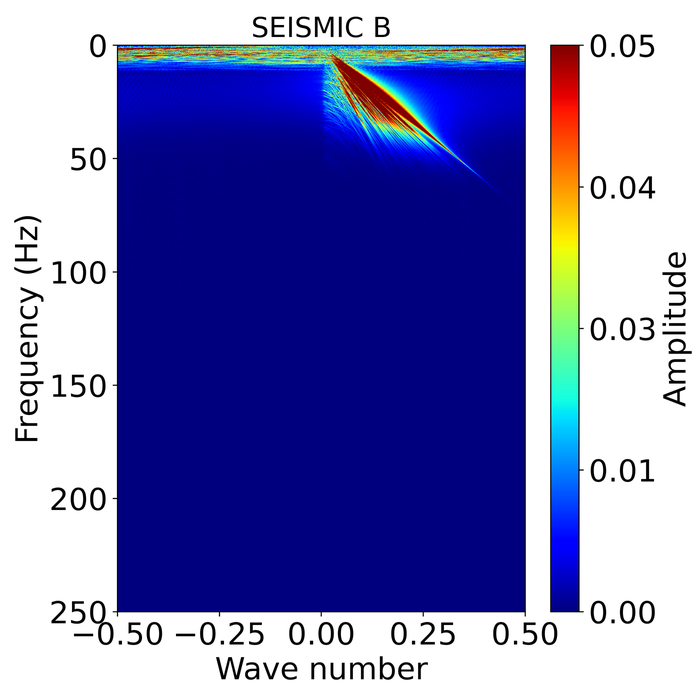}
    \end{subfigure}
    \hfill
    \begin{subfigure}[b]{0.24\textwidth}
        \includegraphics[width=\textwidth]{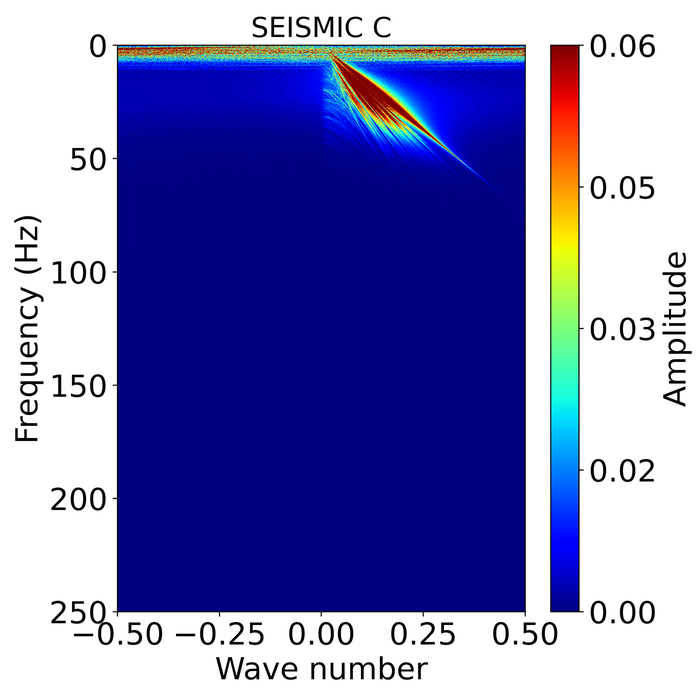}
    \end{subfigure}
    \hfill
    \begin{subfigure}[b]{0.24\textwidth}
        \includegraphics[width=\textwidth]{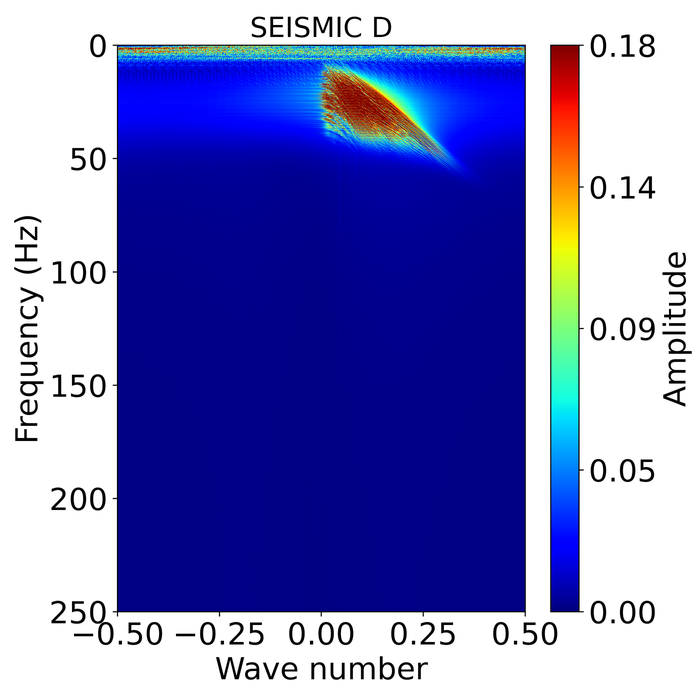}
    \end{subfigure}
    
    \vskip\baselineskip
    
    \begin{subfigure}[b]{0.24\textwidth}
        \includegraphics[width=\textwidth]{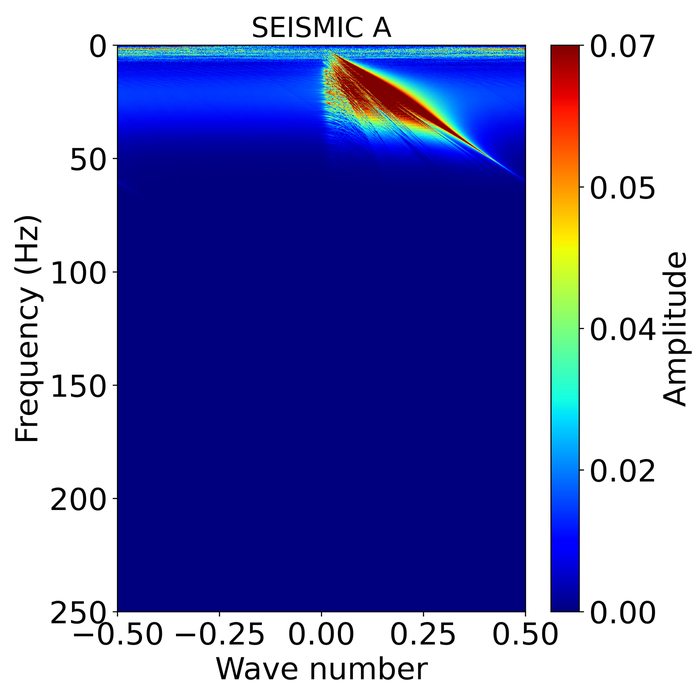}
    \end{subfigure}
    \hfill
    \begin{subfigure}[b]{0.24\textwidth}
        \includegraphics[width=\textwidth]{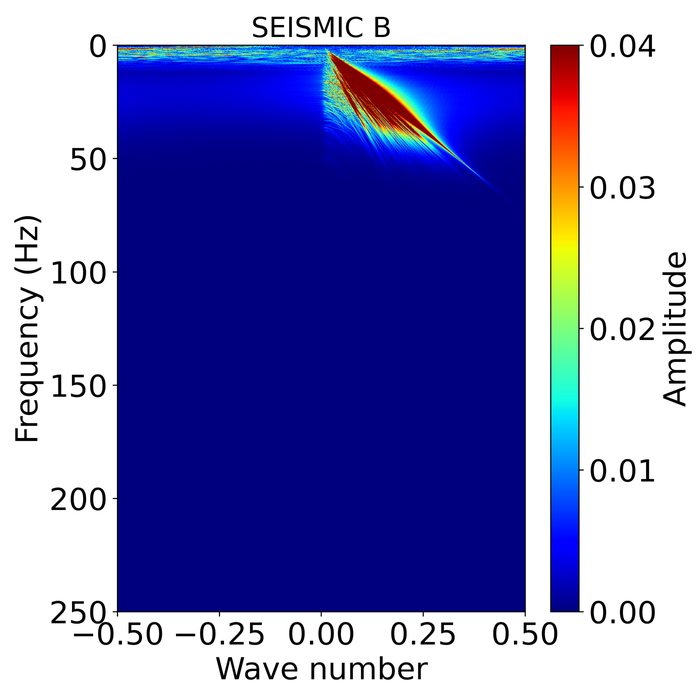}
    \end{subfigure}
    \hfill
    \begin{subfigure}[b]{0.24\textwidth}
        \includegraphics[width=\textwidth]{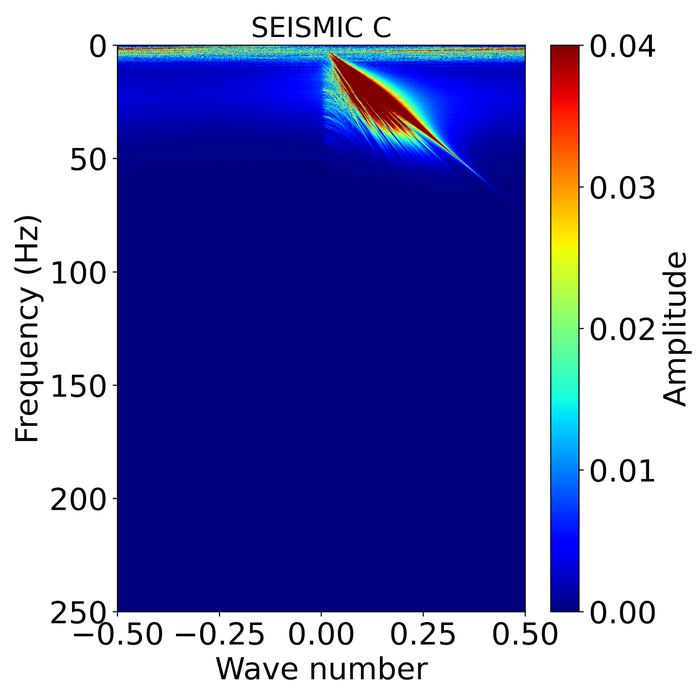}
    \end{subfigure}
    \hfill
    \begin{subfigure}[b]{0.24\textwidth}
        \includegraphics[width=\textwidth]{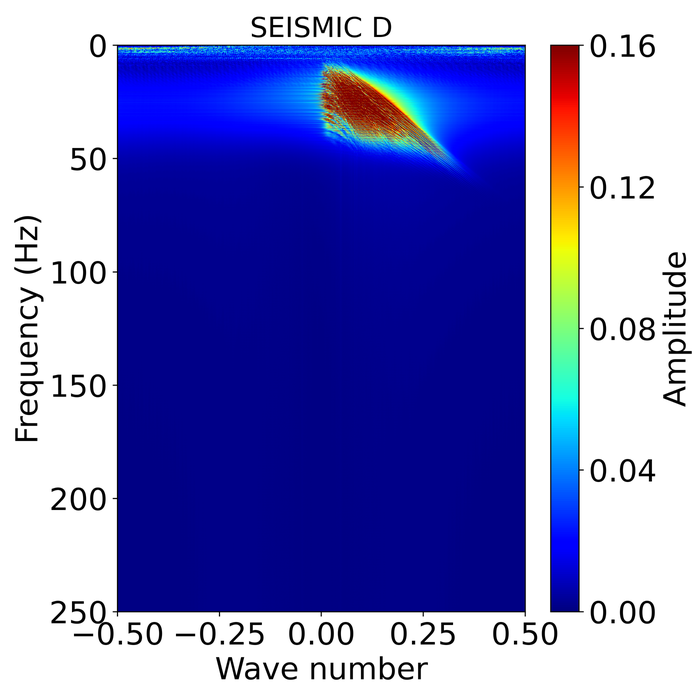}
    \end{subfigure}
    
    \vskip\baselineskip
    
    \begin{subfigure}[b]{0.24\textwidth}
        \includegraphics[width=\textwidth]{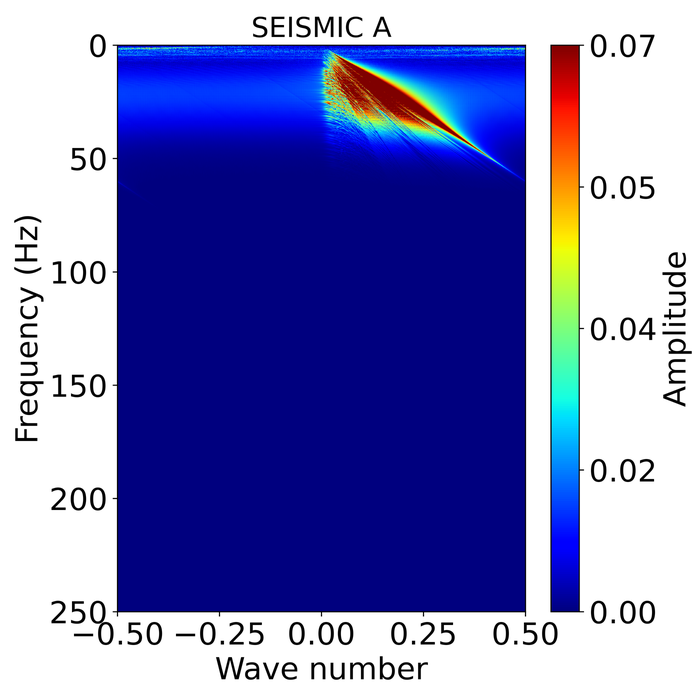}
    \end{subfigure}
    \hfill
    \begin{subfigure}[b]{0.24\textwidth}
        \includegraphics[width=\textwidth]{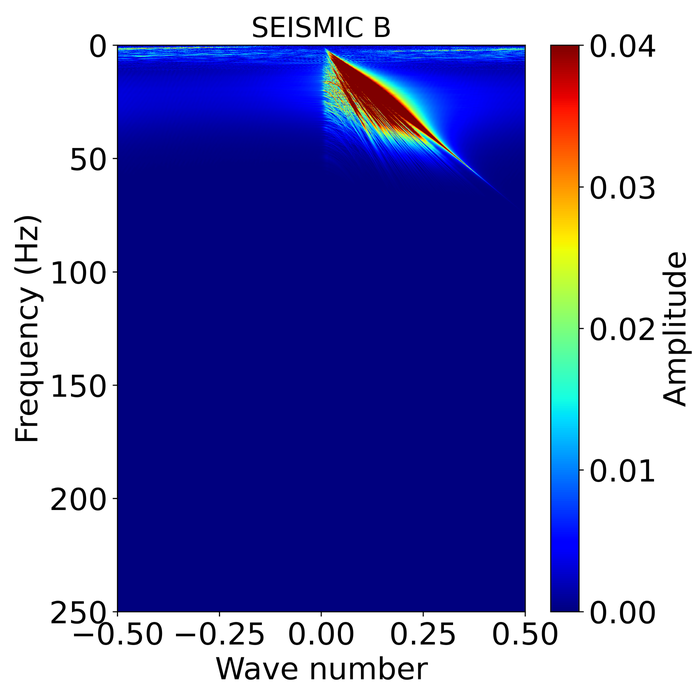}
    \end{subfigure}
    \hfill
    \begin{subfigure}[b]{0.24\textwidth}
        \includegraphics[width=\textwidth]{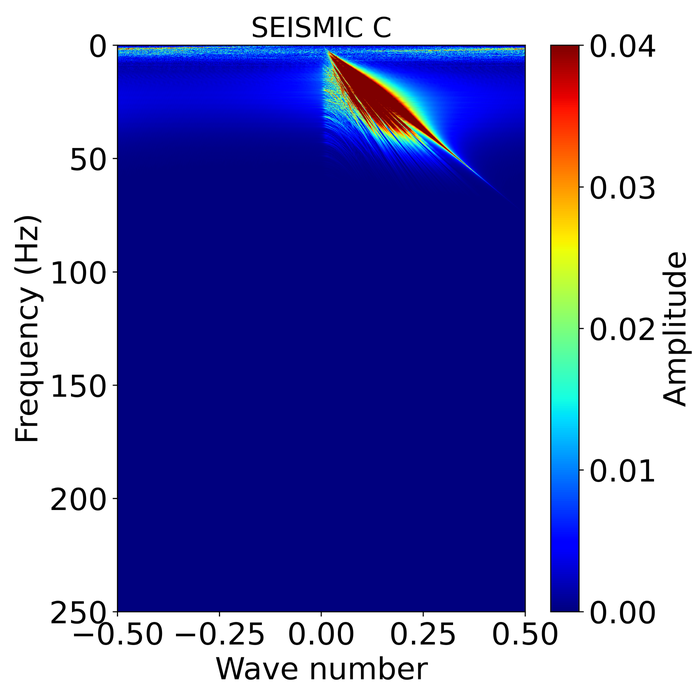}
    \end{subfigure}
    \hfill
    \begin{subfigure}[b]{0.24\textwidth}
        \includegraphics[width=\textwidth]{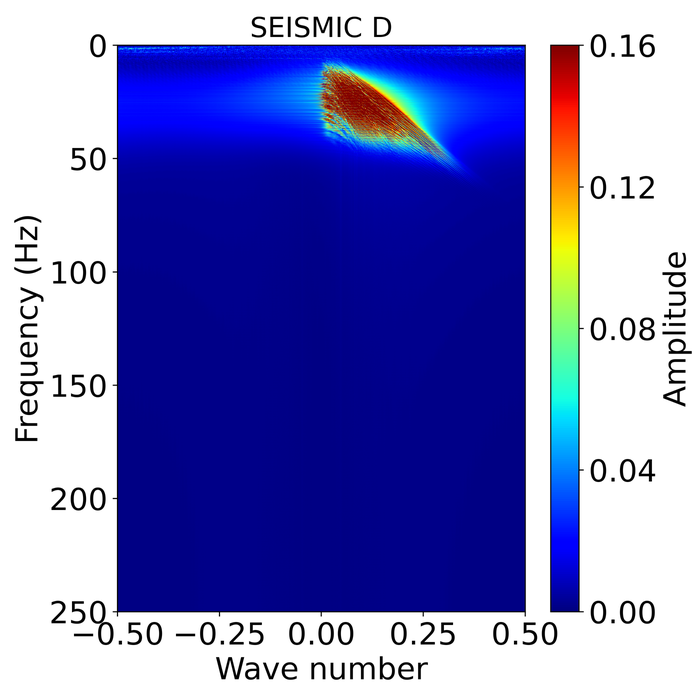}
    \end{subfigure}

    \caption{F-K spectra of the noisy data corrupted with NOISE 1: the columns represent the SEISMIC A, B, C and D data, and the lines represent the noise levels L1, L2, L5 and L10.}
    \label{fig:A1}
\end{figure}

\begin{figure}[ht]
    \centering
    \begin{subfigure}[b]{0.24\textwidth}
        \includegraphics[width=\textwidth]{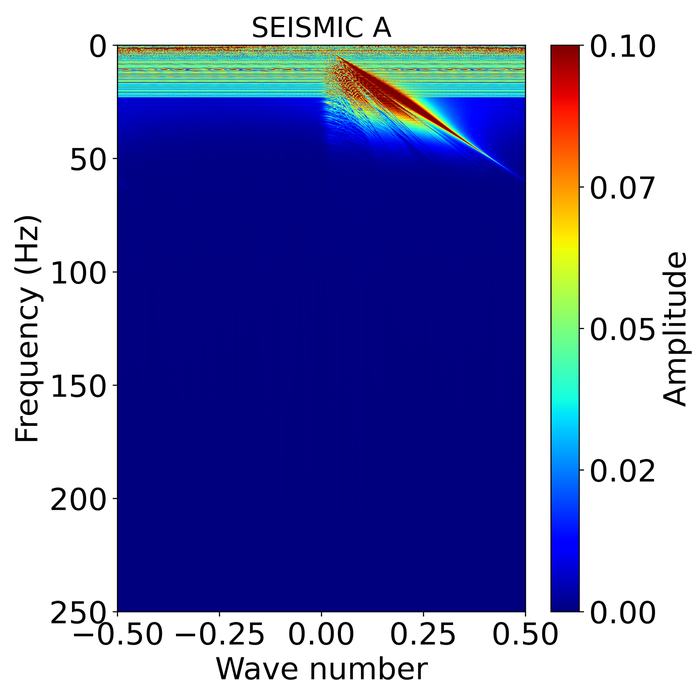}
    \end{subfigure}
    \hfill
    \begin{subfigure}[b]{0.24\textwidth}
        \includegraphics[width=\textwidth]{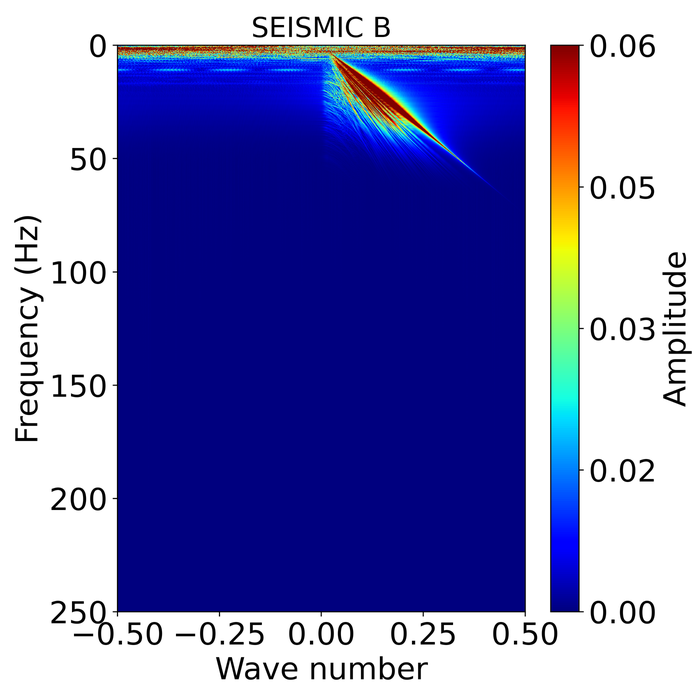}
    \end{subfigure}
    \hfill
    \begin{subfigure}[b]{0.24\textwidth}
        \includegraphics[width=\textwidth]{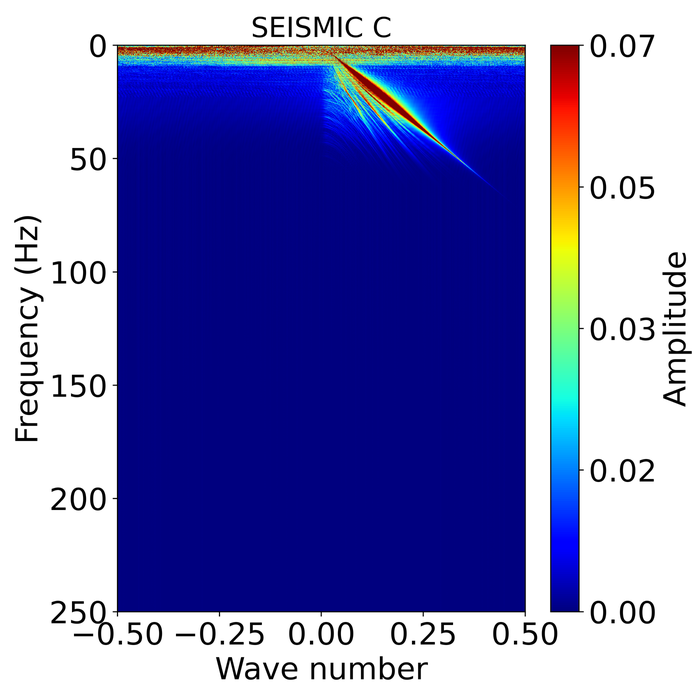}
    \end{subfigure}
    \hfill
    \begin{subfigure}[b]{0.24\textwidth}
        \includegraphics[width=\textwidth]{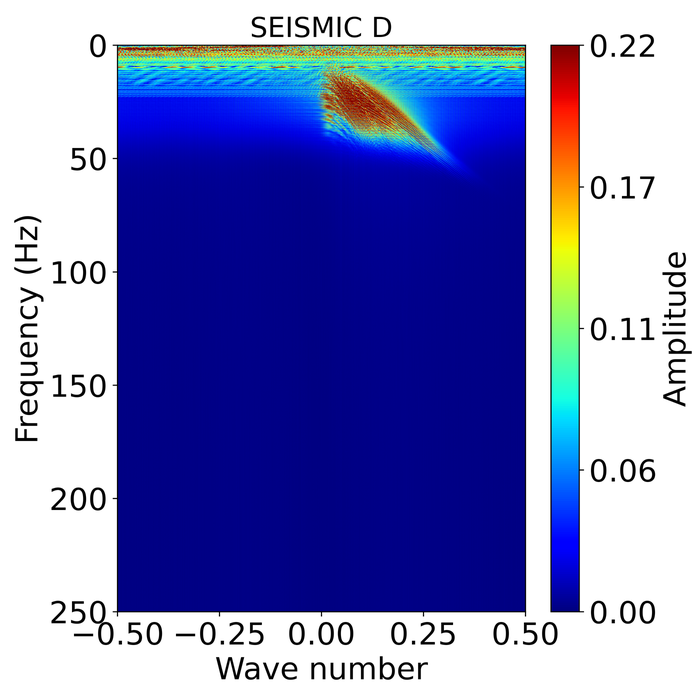}
    \end{subfigure}
    
    \vskip\baselineskip
    
    \begin{subfigure}[b]{0.24\textwidth}
        \includegraphics[width=\textwidth]{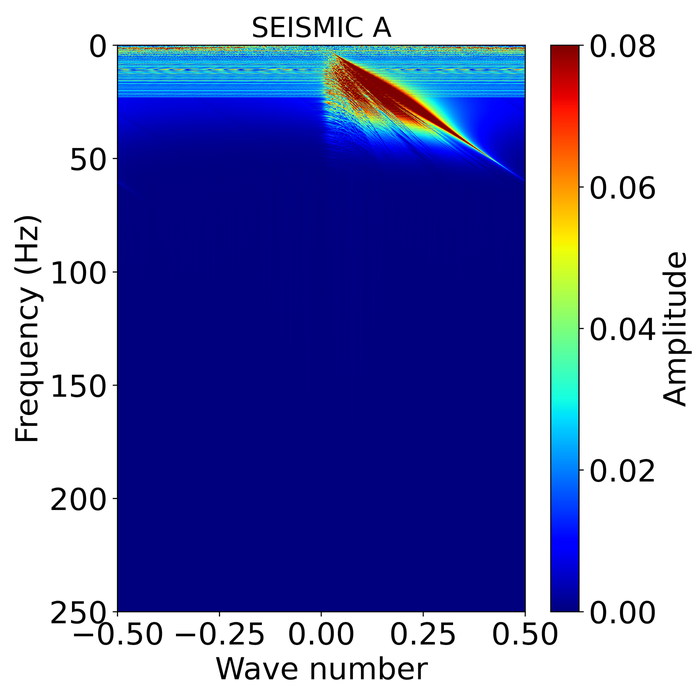}
    \end{subfigure}
    \hfill
    \begin{subfigure}[b]{0.24\textwidth}
        \includegraphics[width=\textwidth]{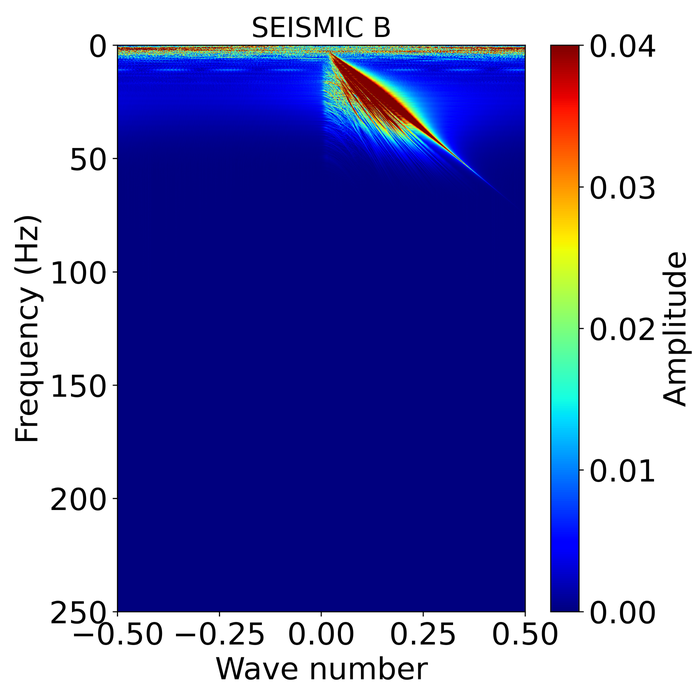}
    \end{subfigure}
    \hfill
    \begin{subfigure}[b]{0.24\textwidth}
        \includegraphics[width=\textwidth]{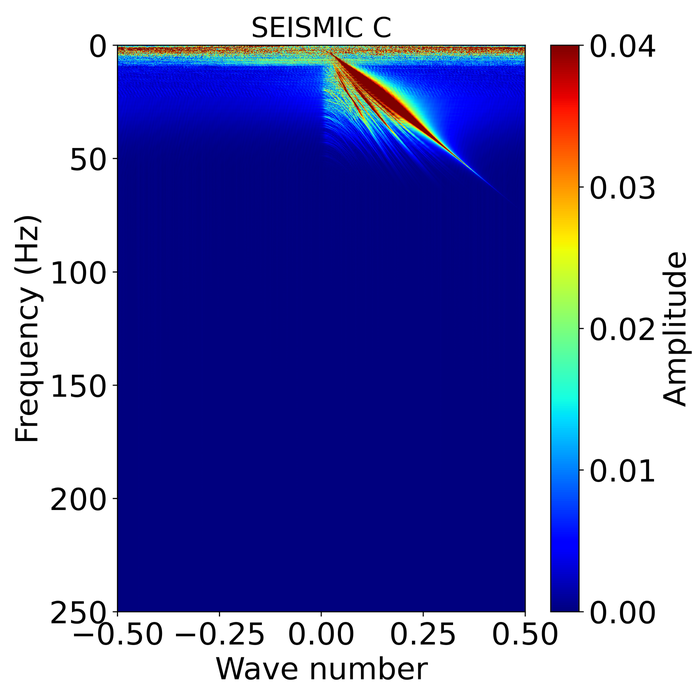}
    \end{subfigure}
    \hfill
    \begin{subfigure}[b]{0.24\textwidth}
        \includegraphics[width=\textwidth]{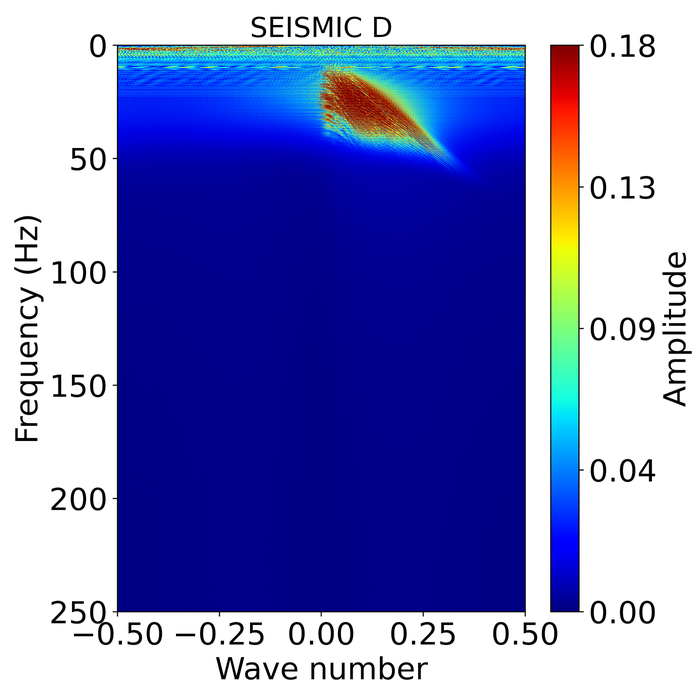}
    \end{subfigure}
    
    \vskip\baselineskip
    
    \begin{subfigure}[b]{0.24\textwidth}
        \includegraphics[width=\textwidth]{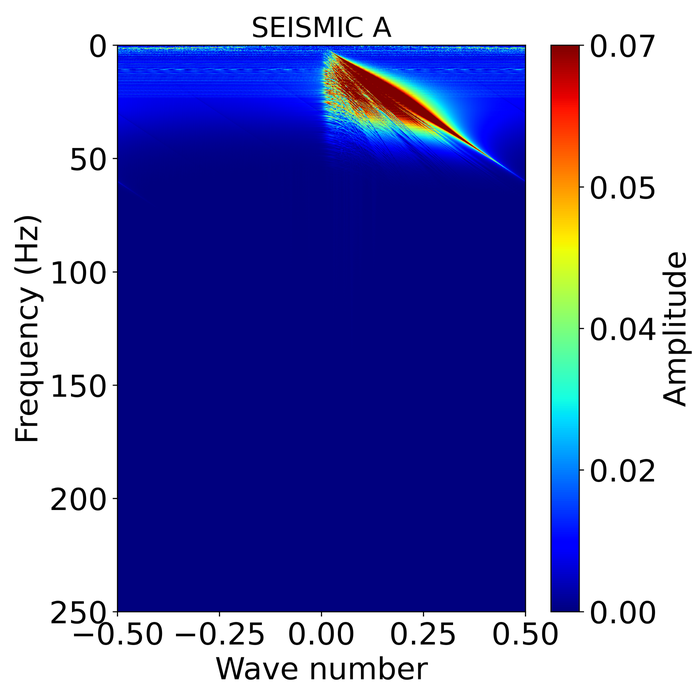}
    \end{subfigure}
    \hfill
    \begin{subfigure}[b]{0.24\textwidth}
        \includegraphics[width=\textwidth]{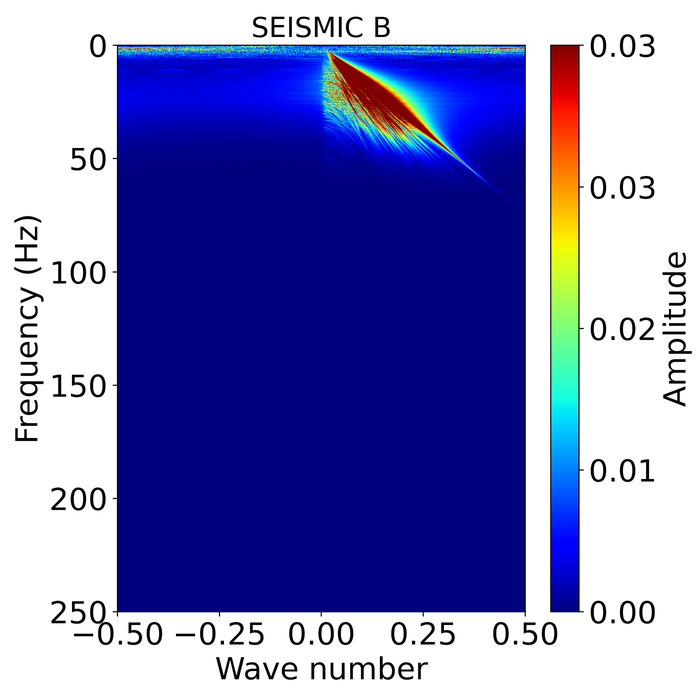}
    \end{subfigure}
    \hfill
    \begin{subfigure}[b]{0.24\textwidth}
        \includegraphics[width=\textwidth]{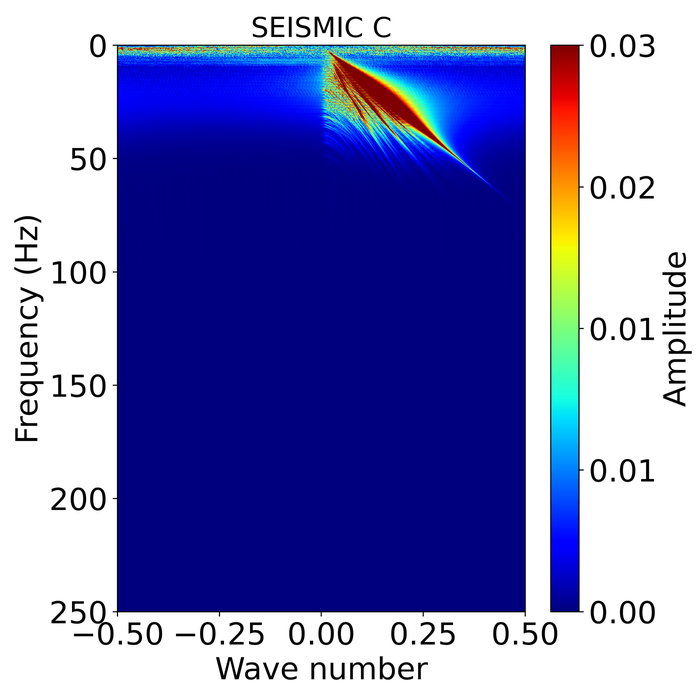}
    \end{subfigure}
    \hfill
    \begin{subfigure}[b]{0.24\textwidth}
        \includegraphics[width=\textwidth]{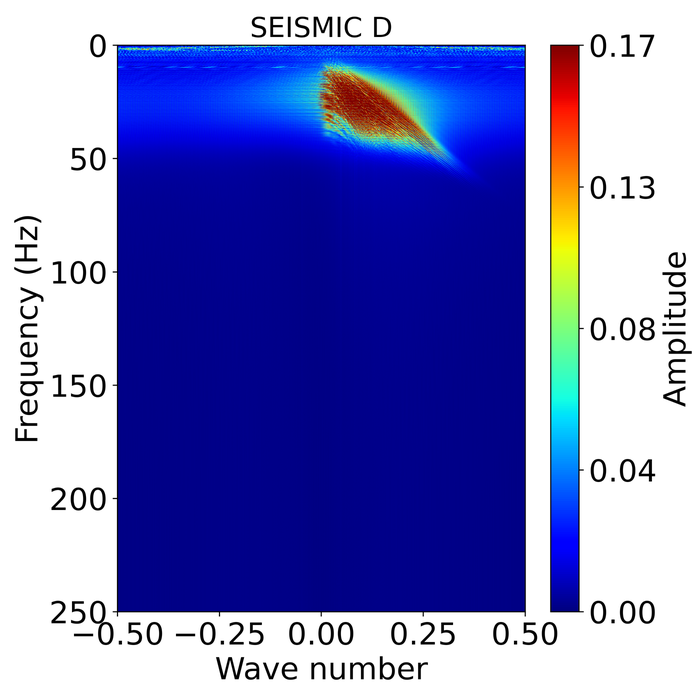}
    \end{subfigure}
    
    \vskip\baselineskip
    
    \begin{subfigure}[b]{0.24\textwidth}
        \includegraphics[width=\textwidth]{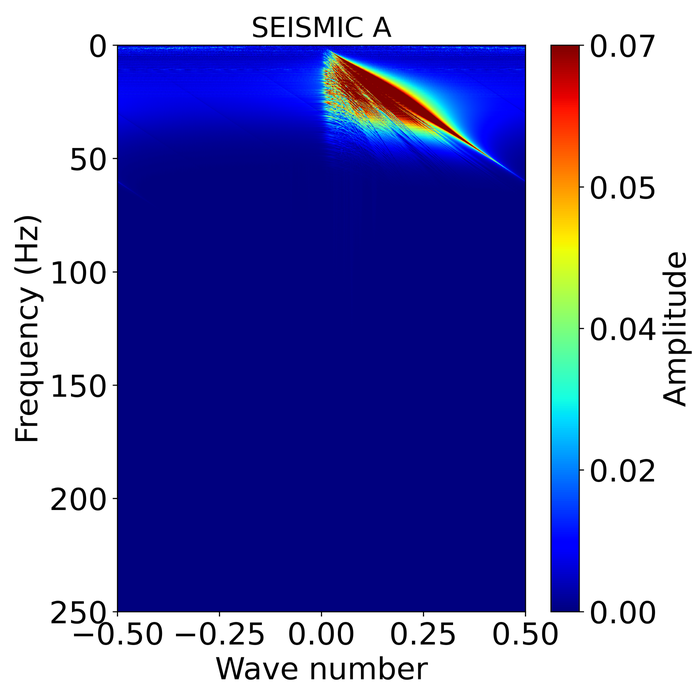}
    \end{subfigure}
    \hfill
    \begin{subfigure}[b]{0.24\textwidth}
        \includegraphics[width=\textwidth]{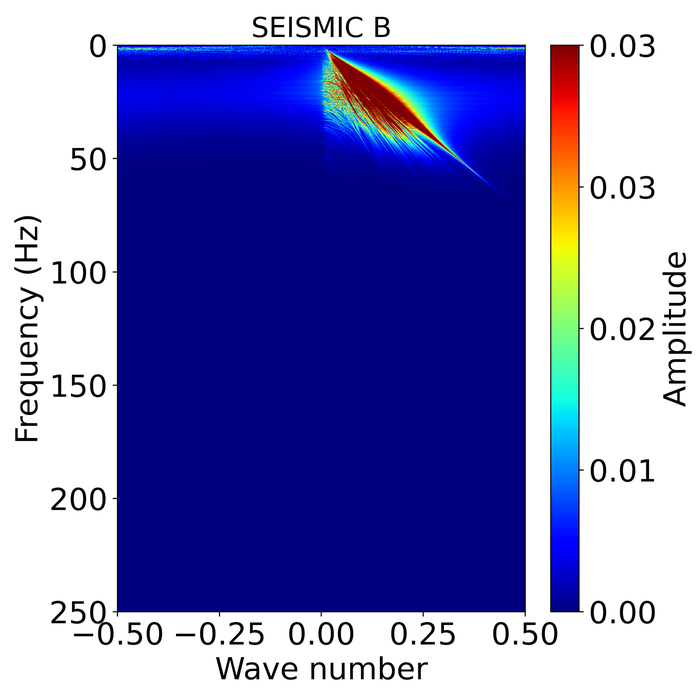}
    \end{subfigure}
    \hfill
    \begin{subfigure}[b]{0.24\textwidth}
        \includegraphics[width=\textwidth]{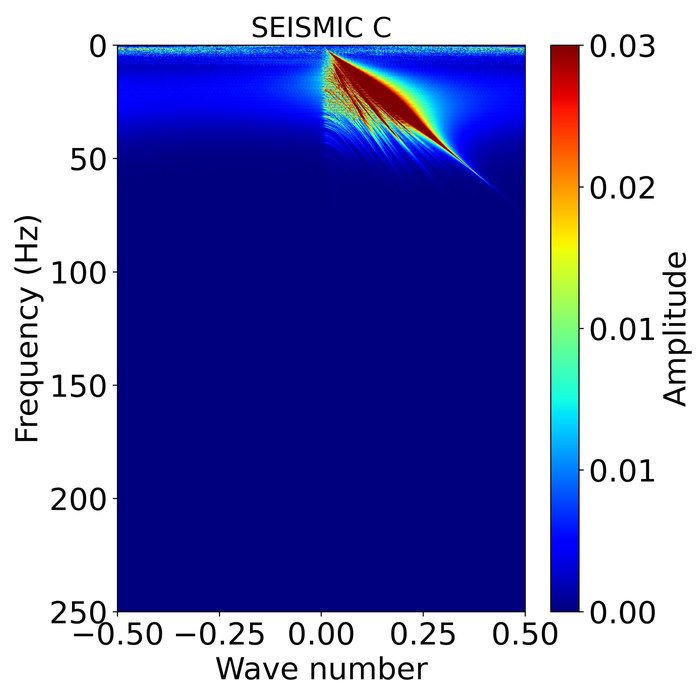}
    \end{subfigure}
    \hfill
    \begin{subfigure}[b]{0.24\textwidth}
        \includegraphics[width=\textwidth]{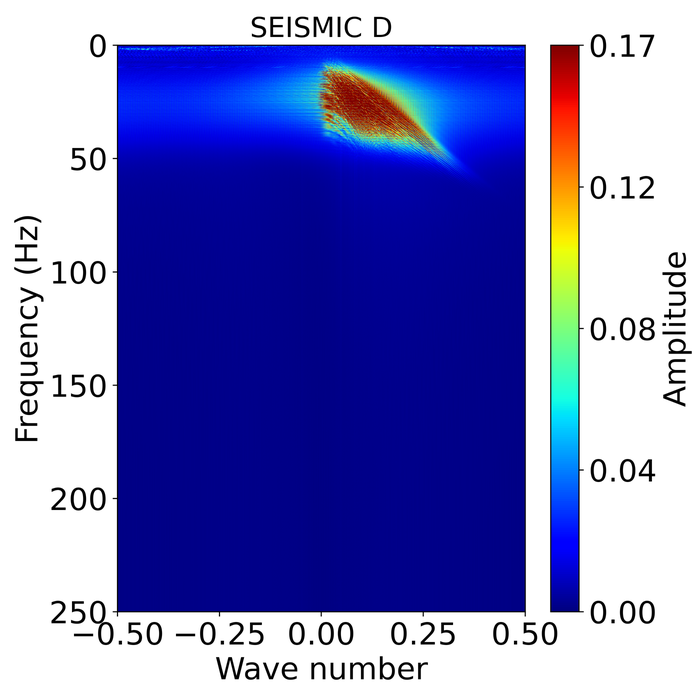}
    \end{subfigure}

    \caption{F-K spectra of the noisy data corrupted with NOISE 2: the columns represent the SEISMIC A, B, C and D data, and the lines represent the noise levels L1, L2, L5 and L10.}
    \label{fig:A2}
\end{figure}

\begin{table}[!htpb]
\centering
\caption{Training and testing sets for model evaluation purposes.}
\label{tab:tableA1}
\begin{tabular}{lll lll}
\hline\noalign{\smallskip}
\textbf{SEISMIC} & \textbf{NOISE} & \textbf{LEVEL} & \textbf{SEISMIC} & \textbf{NOISE} & \textbf{LEVEL} \\
\noalign{\smallskip}\hline\noalign{\smallskip}
\multicolumn{3}{l}{\textbf{TRAINING}} & \multicolumn{3}{l}{\textbf{TESTING}} \\
\noalign{\smallskip}\hline\noalign{\smallskip}
SEISMIC B & NOISE 2 & L2 L5 L10 & SEISMIC A & NOISE 1 & L1 \\
SEISMIC C & NOISE 1 & L2 L5 L10 & SEISMIC D & NOISE 2 & L1 \\
SEISMIC C & NOISE 2 & L2 L5 L10 & SEISMIC D & NOISE 1 & L1 \\
SEISMIC D & NOISE 1 & L2 L5 L10 & SEISMIC C & NOISE 2 & L1 \\
SEISMIC D & NOISE 2 & L2 L5 L10 & SEISMIC C & NOISE 1 & L1 \\
SEISMIC A & NOISE 1 & L1 L5 L10 & SEISMIC B & NOISE 2 & L2 \\
SEISMIC A & NOISE 2 & L1 L5 L10 & SEISMIC B & NOISE 1 & L2 \\
SEISMIC B & NOISE 1 & L1 L5 L10 & SEISMIC A & NOISE 2 & L2 \\
SEISMIC B & NOISE 2 & L1 L5 L10 & SEISMIC A & NOISE 1 & L2 \\
SEISMIC C & NOISE 1 & L1 L5 L10 & SEISMIC D & NOISE 2 & L2 \\
SEISMIC C & NOISE 2 & L1 L5 L10 & SEISMIC D & NOISE 1 & L2 \\
SEISMIC D & NOISE 1 & L1 L5 L10 & SEISMIC C & NOISE 2 & L2 \\
SEISMIC D & NOISE 2 & L1 L2 L10 & SEISMIC C & NOISE 1 & L5 \\
SEISMIC A & NOISE 1 & L1 L2 L10 & SEISMIC B & NOISE 2 & L5 \\
SEISMIC A & NOISE 2 & L1 L2 L10 & SEISMIC B & NOISE 1 & L5 \\
SEISMIC B & NOISE 1 & L1 L2 L10 & SEISMIC A & NOISE 2 & L5 \\
SEISMIC B & NOISE 2 & L1 L2 L10 & SEISMIC A & NOISE 1 & L5 \\
SEISMIC C & NOISE 1 & L1 L2 L10 & SEISMIC D & NOISE 2 & L5 \\
SEISMIC C & NOISE 2 & L1 L2 L10 & SEISMIC D & NOISE 1 & L5 \\
SEISMIC D & NOISE 1 & L1 L2 L10 & SEISMIC C & NOISE 2 & L5 \\
SEISMIC D & NOISE 2 & L1 L2 L10 & SEISMIC C & NOISE 1 & L5 \\
SEISMIC A & NOISE 1 & L1 L2 L5 & SEISMIC B & NOISE 2 & L10 \\
SEISMIC A & NOISE 2 & L1 L2 L5 & SEISMIC B & NOISE 1 & L10 \\
SEISMIC B & NOISE 1 & L1 L2 L5 & SEISMIC A & NOISE 2 & L10 \\
SEISMIC B & NOISE 2 & L1 L2 L5 & SEISMIC A & NOISE 1 & L10 \\
SEISMIC C & NOISE 1 & L1 L2 L5 & SEISMIC D & NOISE 2 & L10 \\
SEISMIC C & NOISE 2 & L1 L2 L5 & SEISMIC D & NOISE 1 & L10 \\
SEISMIC D & NOISE 1 & L1 L2 L5 & SEISMIC C & NOISE 2 & L10 \\
SEISMIC D & NOISE 2 & L1 L2 L5 & SEISMIC C & NOISE 1 & L10 \\
\noalign{\smallskip}\hline
\end{tabular}
\end{table}
\end{document}